\documentclass[fleqn,12pt,twoside]{article}
\usepackage{espcrc1}

\usepackage{graphicx}
\usepackage[figuresright]{rotating}

\newcommand{\AmS}{{\protect\the\textfont2
  A\kern-.1667em\lower.5ex\hbox{M}\kern-.125emS}}

\title{Approximate coherent states for nonlinear systems}

\author{R. Rom\'an-Ancheyta\address{Instituto de Ciencias F\'isicas,
        Universidad Nacional Aut\'onoma de M\'exico\\ 
        C.P. 62210 Cuernavaca, Morelos, M\'exico}
       and J. R\'ecamier\addressmark }
      
\begin{document}

% typeset front matter
\maketitle
\tableofcontents

\begin{abstract}
On the basis of the f-deformed oscillator formalism, we propose to construct nonlinear coherent states for Hamiltonian systems having linear and quadratic terms in the the number operator by means of the two following definitions: i) as deformed annihilation operator coherent states (AOCS) and ii) as deformed displacement operator coherent states (DOCS). For the particular cases of the Morse and Modified P\"oschl-Teller potentials, modeled as f-deformed oscillators (both supporting a finite number of bound states), the properties of their corresponding nonlinear coherent states, viewed as DOCS, are analyzed in terms of their occupation number distribution, their evolution on phase space, and their uncertainty relations. 
\end{abstract}

  \section{Introduction}
The idea of creating a coherent state for a quantum system was first proposed by Schr\"odinger, in 1926, in connection with the classical states of the quantum harmonic oscillator; he referred to them as states of the minimum uncertainty product \cite{schroedinger}. Almost forty years later in 1963, Glauber \cite{glauber} introduced the field coherent states, these have been realized experimentally with the development of lasers, and can be obtained from any one of three mathematical definitions: (i) as the right-hand eigenstates of the boson annihilation operator $\hat a|\alpha\rangle = \alpha|\alpha\rangle$, with $\alpha$ being a complex number; (ii) as the states obtained by application of the displacement operator $D(\alpha)= \exp(\alpha \hat a^{\dagger}-\alpha^{*}\hat a)$ upon the vacuum state; and (iii) as those states with a minimum uncertainty relationship $(\Delta q)^2(\Delta p)^2 = 1/4$, with $q=(\hat a+\hat a^{\dagger})/\sqrt{2}$ and $p=i(\hat a^{\dagger}-\hat a)/\sqrt{2}$ the position and momentum operators respectively and $\Delta q = \Delta p$. The same states are obtained from any one of the three mathematical definitions when one makes use of the harmonic oscillator algebra. At the same time, Klauder \cite{klauder} developed a set of continuous states in which the basic ideas of coherent states for arbitrary Lie groups were contained. The complete construction of coherent states of Lie groups was achieved by Perelomov \cite{perelomov} and Gilmore \cite{gilmore} almost ten years later; the basic theme of this development was to connect the coherent states with the dynamical group for each physical problem thus allowing the generalization of the concept of coherent state. For systems far from the ground state or with a finite number of bound states, the harmonic oscillator algebra is no longer adequate, therefore there has arisen an interest in generalizing the concept of coherent state for systems with different dynamical properties.
     
Nieto and Simmons \cite{nieto} generalized the notion of coherent states for potentials having unequally spaced energy levels. Their construction is such that the resultant states are localized, follow the classical motion and disperse as little as possible in time. Their coherent states are defined as those which minimize the uncertainty relation equation. They constructed coherent states for the P\"oschl-Teller potential, the harmonic oscillator with a centripetal barrier and the Morse potential. 
   
   Gazeau and Klauder \cite{gazeau} proposed a generalization for systems with one degree of freedom possessing discrete and continuous spectra. These states present continuity of labeling, resolution of unity, and temporal stability. The key point is to parametrize such states by means of two real values: an amplitude $J$ and a phase $\gamma$, instead of using a complex value $\alpha$.
   
 Man'ko and collaborators \cite{manko}  introduced coherent states of a {\em f-deformed} algebra as eigenstates of a deformed annihilation operator $\hat A= \hat a f(\hat n)$, where $\hat a$ is the usual annihilation operator of the harmonic oscillator algebra and $f(\hat n)$ is a number operator function that specifies the potential. These states present nonclassical properties such as squeezing and anti-bunching. More recently, they generalized the displacement operator method for the case of {\em f-deformed} oscillators introducing a deformed version of the displacement operator \cite{manko2} with the disadvantage that it is non unitary and does not displace the deformed operators in the usual form.
 
On the basis of the f-deformed oscillator formalism, in previous works we put forward the construction of generalized coherent states for nonlinear systems by selecting a specific deformation function in a way such that the energy spectrum of the Hamiltonian it seeks to describe is similar to that of a f-deformed Hamiltonian. In this regard, we examined the trigonometric and the modified P\"oschl-Teller potentials, as well as the Morse potential, all of them containing linear and quadratic terms in the number operator \cite{santos11}. In addition to these, a harmonic oscillator with an inverse square potential in two dimensions  was also considered \cite{ricardo}. 
 
Coherent states have been applied not only in optics but in many fields of physics, for instance, in the study of the forced harmonic oscillator \cite{carru}, quantum oscillators \cite{glauberpl}, quantum nondemolition measurements \cite{unru}, nonequilibrium statistical mechanics\cite{taka}, and atomic and molecular physics \cite{arrechi,wang,advq}. Excellent reviews can be found in \cite{klauderb,ali,gilmore2}. \\
 
 The paper is organized as follows. In section \ref{Harmonic} we briefly describe the properties of the usual coherent states. In section \ref{Deformed} we present a methodology to construct deformed Hamiltonians for specific systems. In section \ref{NLCS} we give two alternative forms to obtain the coherent states pertinent to a nonlinear Hamiltonian. And finally, in section \ref{RES} we present some numerical results.

  \section{Harmonic oscillator or field coherent states}\label{Harmonic}

The harmonic oscillator coherent states, also called {\it field coherent states} \cite{glauber}, are quantum states of minimum uncertainty product which most closely resemble the classical ones in the sense that they remain well localized around their corresponding classical trajectory. When one makes use of the harmonic oscillator algebra, the same coherent states are obtained from the three Glauber's mathematical definitions mentioned above. Such states take the following form in terms of the number state basis:

  \begin{equation}
  |\alpha\rangle = e^{-\frac{1}{2}|\alpha|^2}\sum_{n=0}^{\infty} \frac{\alpha^n}{\sqrt{n!}}|n\rangle,
  \end{equation}
  where $\alpha$ is an arbitrary complex number. For every complex number $\alpha\neq 0$ the coherent state $|\alpha\rangle$ has a non zero projection on every Fock state $|n\rangle$.
  
From a statistical point of view, it follows from the above that the occupation  number distribution of a coherent state, $P_{\alpha}(n)=|\langle n|\alpha\rangle|^2$, is characterized by a Poisson distribution:
\begin{equation}
P_{\alpha}(n) = e^{-|\alpha|^2}\frac{|\alpha|^{2n}}{n!}, 
\end{equation}
  with an average occupation number $\bar n=|\alpha|^2$ and mean square root deviation 
  \[ \Delta N = \sqrt{\langle \alpha|N^2|\alpha\rangle -\langle \alpha|N|\alpha\rangle^2} = |\alpha|=\sqrt{\bar n}. \]
Such states are not orthogonal, and the overlap between two of them is given by  
\begin{equation}
\langle \alpha|\beta\rangle = e^{-\frac{1}{2}|\beta-\alpha|^2}e^{\frac{1}{2}(\beta^*\alpha-\beta\alpha^*)}
\end{equation}
Moreover, they form an over-complete basis with the closure relationship,
\begin{equation}
 {\cal I} = \frac{1}{\pi}\int |\alpha\rangle \langle \alpha| d^2\alpha ,
 \end{equation}
  so that any state of the group generated by the operators $\hat a$, $\hat a^{\dagger}$ and $\hat n=\hat a^{\dagger}\hat a$ can be expanded in terms of coherent states. 
  
  \section{Deformed Oscillators}\label{Deformed}
  A deformed oscillator is a non-harmonic system characterized by a Hamiltonian of the harmonic oscillator form 
\begin{equation}  
\hat H_D=\frac{\hbar \Omega}{2}\left(\hat A\hat A^{\dagger} + \hat A^{\dagger}\hat A\right), 
\end{equation}
  with a specific frequency $\Omega$ and deformed boson creation $\hat A^{\dagger}$ and annihilation $\hat A$ operators defined as \cite{manko}: 
\begin{equation}
 \hat A^{\dagger} = f(\hat n)\hat a^{\dagger} = \hat a^{\dagger}f(\hat n+1), \ \ \ \hat A =\hat a f(\hat n) = f(\hat n+1)\hat a ,
 \end{equation}
  with $f(\hat n)$ a number operator function that specifies the deformation. The commutation relations between the deformed operators are:
  \begin{equation}
  [\hat A, \hat A^{\dagger}] = (\hat n+1)f^2(\hat n+1)-\hat n f^2(\hat n), \ \ [\hat A, \hat n]=\hat A, \ \ [\hat A^{\dagger},\hat n] = -\hat A^{\dagger}.
  \end{equation}
  Notice that the commutation relations between the deformed operators $\hat A$,\ $\hat A^{\dagger}$ and the number operator are the same as those among the usual operators $\hat a$,\ $\hat a^{\dagger}$ with the number operator.  Note that the eigenfunctions of the harmonic oscillator $|n\rangle$ are also eigenfunctions of the deformed oscillator. In terms of the number operator, the deformed Hamiltonian takes the form:
  \begin{equation}\label{eq:Hdef}
  \hat H_D = \frac{\hbar\Omega}{2}\left(\hat n f^2(\hat n)+(\hat n+1)f^2(\hat n+1)\right).
  \end{equation}
  \subsection{Algebraic Hamiltonian for the Morse potential}
  The Morse oscillator is a particularly useful anharmonic potential  for the description of systems that deviate from the ideal harmonic oscillator conduct and has been used widely to model the vibrations of a diatomic molecule. 
  If we choose the deformation function \cite{gorayeb}:
  \begin{equation}\label{eq:fdefmorse}
  f^2(\hat n) = 1-\chi_a \hat n,
  \end{equation}
  with $\chi_a$ an anharmonicity parameter, and substitute it into Eq.~\ref{eq:Hdef}, we obtain an algebraic Hamiltonian of the form
  \begin{equation}
 \hat H_D= \hbar\Omega\left[ \hat n+\frac{1}{2}-\chi_a\left(\hat n+\frac{1}{2}\right)^2 -\frac{\chi_a}{4}\right],
  \end{equation}
  whose spectrum is in essence identical to that of the Morse potential \cite{landau}, i.e, 
  \begin{equation}
  E_n=\hbar\omega_e\left(n+\frac{1}{2}\right)-\frac{\hbar\omega_e}{2N+1}\left(n+\frac{1}{2}\right)^2,
  \end{equation}
provided that $\omega_e = \Omega$ and $\chi_a = 1/(2N+1)$, with $N$ being the number of bound states corresponding to the integers $0\leq n\leq N-1$. For this particular choice of deformation function, the commutator between deformed operators is:
  \begin{equation}
  [\hat A,\hat A^{\dagger}] = 1-\chi_a(2\hat n+1) = \frac{2(N-\hat{n})}{2N+1},
  \end{equation}
from which one can see that the commutator equals a scalar plus a linear function of the number operator.\\  

  The action of the deformed operators upon the number states is:
  \begin{equation}
  \hat A|n\rangle = \sqrt{n}f(n)|n-1\rangle = \sqrt{n(1-\chi_a n)}|n-1\rangle 
  \end{equation}
  and
  \begin{equation}
  \hat A^{\dagger}|n\rangle = \sqrt{n+1}f(n+1)|n+1\rangle = \sqrt{(n+1)(1-\chi_a(n+1))}|n+1\rangle.
  \end{equation}
  The deformed operators change the number of quanta in $\pm 1$ and their corresponding matrix elements depend on the deformation function $f(n)$. Furthermore, from the commutation relations we see that the set $\{\hat A, \hat A^{\dagger}, \hat n, 1\}$ is closed under the operation of commutation.

 \subsection{Algebraic Hamiltonian for the Modified P\"oschl-Teller potential}
    The modified P\"oschl-Teller potential can be written as:
  \begin{equation}
  V(x)= U_0 \tanh^2(ax)
  \end{equation}  
    where $U_0$ is the depth of the well, $a$ is the range of the potential and $x$ is the relative distance from the equilibrium position. The  eigenfunctions and eigenvalues are, respectively, \cite{landau}
    \begin{equation}
    \psi_{n}^{\epsilon}(\zeta) = N_{n}^{\epsilon}(1-\zeta^2)^{\epsilon/2}F(-n,\epsilon+s+1;\epsilon+1,(1-\zeta)/2)
    \end{equation}
    and
   \begin{equation}
   E_n=U_0 -\frac{\hbar^2a^2}{8\mu}\left(-(2n+1)+\sqrt{1+\frac{8\mu U_0}{\hbar^2 a^2}}\right)^2,
   \end{equation}
   where $N_{n}^{\epsilon}$ is a normalization constant, $\zeta=\tanh(ax)$, $\mu$ is the reduced mass of the molecule, $s$ is related with the depth of the well so that $s(s+1)=2\mu U_0/(\hbar^2 a^2)$, $\epsilon=\sqrt{-2\mu (E-U_0)}/\hbar a$ and $F(a,b;c,z)$ stands for the hypergeometric function \cite{abramowitz}. If we write the eigenvalues in terms of the parameter $s$, we obtain

   \begin{equation}\label{eq:modpt}
   E_n = \frac{\hbar^2 a^2}{2\mu}\left(s+2sn -n^2\right).
   \end{equation}
   The number of bound states is determined by the dissociation limit $\epsilon=s-n_{max}=0.$ For integer values of $s$, the state associated with null energy is not normalizable, whereby the last bound state corresponds to $n_{max}=s-1$ \cite{lemus}.  The modified P\"oschl-Teller potential is a nonlinear potential symmetric in the displacement coordinate. \\
   
 Let us now consider a deformation function of the form \cite{santos11}
   \begin{equation}
   f^2(\hat n)= \frac{\hbar a^2}{2\mu\Omega}\left(2s+1-\hat n\right).
   \end{equation}
 By substituting it into Eq.~\ref{eq:Hdef}, we obtain the deformed Hamiltonian
   \begin{equation}
   \hat H_D= \frac{\hbar^2 a^2}{2\mu}\left(-\hat n^2+2s\hat n +s\right),
   \end{equation}
   whose spectrum is identical to that of Eq.~\ref{eq:modpt}. The harmonic limit is obtained by taking $s\rightarrow\infty$, $a\rightarrow 0$ with $sa^2\rightarrow \mu\Omega/\hbar$.
   
   For this choice of deformation function, deformed operators, together with the number operator, obey the following commutation relations:
   \begin{equation}
   [\hat A, \hat A^{\dagger}] = \frac{\hbar a^2}{\mu\Omega}(s-\hat n), \ \ [\hat A, \hat n]=\hat A, \ \ [\hat A^{\dagger},\hat n]=-\hat A^{\dagger},
   \end{equation}
   which have the correct harmonic limit and are similar to those of the generators of the $SU(2)$ group.

  \subsection{Algebraic Hamiltonian for the trigonometric P\"oschl-Teller potential}
  The trigonometric P\"oschl-Teller potential is given by:
  \begin{equation}
  V(x) = U_0 \tan^{2}(ax),
   \end{equation}
   where $U_0$ is the strength of the potential and $a$ is its range. This potential supports an infinite number of bound states, its eigenfunctions and eigenvalues are \cite{nieto}:
   \begin{equation}
   \psi_{n}^{\lambda}(x) = \sqrt{\frac{a(\lambda+n)\Gamma(2\lambda+n)}{\Gamma(n+1)}} (\cos(ax))^{1/2} P_{n+\lambda-1/2}^{1/2-\lambda}(\sin(ax)),
   \end{equation}
   \begin{equation}\label{eq:trigpt}
   E_n=\frac{\hbar^2a^2}{2\mu}(n^2+2\lambda n+\lambda)=\hbar\omega\left(n+\frac{1}{2}+\frac{n^2}{2\lambda}\right),
   \end{equation}
   where $\mu$ is the mass of the particle, $\omega=\hbar\lambda a^2/\mu$ and the parameter $\lambda$ is related to the potential strength and range by $\lambda(\lambda-1)=2\mu U_0/\hbar^2 a^2$. In the harmonic limit $\lambda\rightarrow\infty$ and $a\rightarrow 0$ with $\lambda a^2=\mu\omega/\hbar$.
   
   If we choose a deformation function \cite{ricardo}
   \begin{equation}
   f^2(\hat n)=\frac{\hbar a^2}{2\mu\Omega}(\hat n+2\lambda-1),
   \end{equation}
    the deformed Hamiltonian becomes
    \begin{equation}
   \hat H_D= \frac{\hbar^2 a^2}{2\mu}\left(\hat n^2+2\lambda \hat n+\lambda\right),
    \end{equation}
   whose eigenvalues are identical to those given by Eq.~\ref{eq:trigpt}. Once we have given the deformation function, the deformed operators are specified. In this case the commutation relations are:
   \begin{equation}
   [\hat A, \hat A^{\dagger}] = \frac{\hbar a^2}{\mu\Omega}(\hat n+\lambda), \ \ [\hat A, \hat n]=\hat A, \ \ [\hat A^{\dagger},\hat n] = -\hat A^{\dagger},
   \end{equation}
   which are similar to those of the generators of the $SU(1,1)$ group. We emphasize here that  the set of operators $\{\hat A, \hat A^{\dagger},\hat n, 1\}$ is closed under the operation of commutation.
   
   \section{Nonlinear Coherent States}\label{NLCS}
   In this work we consider the generalization of two of the known definitions given to construct the field coherent states, namely, i) as eigenstates of an annihilation operator and ii) as the states obtained from the application of the displacement operator upon a maximal state.

   \subsection{Coherent states as eigenstates of the deformed annihilation operator}
   Following Man'ko and collaborators, we construct deformed coherent states as eigenstates of the deformed annihilation operator \cite{manko}:
   \begin{equation}\label{eq:aocs}
   \hat A|\alpha,f\rangle = \alpha|\alpha,f\rangle.
   \end{equation}
In order to solve Eq. (\ref{eq:aocs}), let the state $|\alpha,f\rangle$ be written as a weighted superposition of the number eigenstates $\{|0\rangle,|1\rangle,\dots,|n\rangle,\dots\}$: %resulting states acquire the explicit form
      \begin{equation}
      |\alpha,f\rangle = N_f \sum_{n=0}^{\infty} c_{n}^{f} |n\rangle.
      \end{equation}
On inserting this into Eq.~\ref{eq:aocs}, we get the following relation between the coefficients $c_{n}^{f}$ and $c_{n-1}^{f}$:
 \begin{equation}
 c_{n}^{f}f(n)\sqrt{n}=\alpha c_{n-1}^{f}.
 \end{equation}
 By applying $n$ times the annihilation operator, we get a relationship between  $c_{n}^{f}$ and $c_{0}^{f}$:
      \begin{equation}\label{eq:recurr}
      c_{n}^{f}f(n)!\sqrt{n!} = \alpha^n c_{0}^{f},
      \end{equation}      
where $f(n)!=f(n)f(n-1)\cdots f(0)$.
  Substitution of Eq.~\ref{eq:recurr} into Eq.~\ref{eq:aocs} gives us the following expression for the Annihilation Operator Coherent States (AOCS):
     \begin{equation}
     |\alpha,f\rangle = N_f \sum_{n=0}^{\infty} \frac{\alpha^n}{\sqrt{n!}f(n)!}|n\rangle.
     \end{equation}
As an example we replaced the explicit form of the deformation function for the trigonometric
P\"oschl-Teller potential and obtain the AOCS associated with this system
	 \begin{equation}\label{AOCS-trigo}
	 |\alpha,f\rangle=N_f\sum_{n=0}^\infty\alpha^n\sqrt{\frac{(2\lambda)^n\Gamma(2\lambda)}
	 {n!\Gamma(2\lambda+n)}}|n\rangle
	 \end{equation}
with $N_f=[_0F_1(2\lambda;2\lambda|\alpha|^2)]^{-1/2}$ a normalization constant.
     
\subsection{Coherent states obtained via the deformed displacement operator}
      In this subsection we construct the Displaced Operator Coherent States (DOCS) for the nonlinear potentials discussed in section \ref{Deformed}. We propose to construct them by application of a deformed displacement operator on the fundamental state of the system, i.e,
      \begin{equation}
      |\zeta(\alpha)\rangle = \exp[\alpha\hat A^{\dagger}-\alpha^{*}\hat A]|0\rangle
      \label{eq:docsdef}
      \end{equation}
      where $\alpha$ is a complex parameter. The deformed displacement operator is obtained from the usual one through the replacement of the harmonic oscillator creation and annihilation operators by their deformed counterparts. Since the commutator between the deformed operators can be a rather complicated function of the number operator, it is not possible, in general, to disentangle the exponential in (\ref{eq:docsdef}). However, for the cases we are considering in this work, we have found that the commutator between the deformed operators $\hat A$, $\hat A^{\dagger}$ is equal to a scalar plus a linear function of the number operator, and therefore the  dynamical group pertinent to these systems is composed by the operators $\{\hat A^{\dagger}, \hat A, \hat n, 1\}$ which form a Lie algebra. So, the deformed displacement operator can be disentangled \cite{puri,gilmorebook} to get (for the particular case of a  Morse potential) \cite{osantos12}
      \begin{eqnarray}
      \hat D_{f}(\alpha)&=&\exp[\alpha\hat A^{\dagger}-\alpha^{*}\hat A], \nonumber \\
      &=& \exp\left(\zeta\frac{\hat A^{\dagger}}{\sqrt{\chi_a}}\right) \left(\frac{1}{1+|\zeta|^2}\right)^{g(\chi_a \hat n)/2\chi_a} \exp\left(-\zeta^{*}\frac{\hat A}{\sqrt{\chi_a}}\right),
      \end{eqnarray}
      where, for a given value of $\alpha= |\alpha|e^{i\phi}$, the complex parameter $\zeta=e^{i\phi}\tan(|\alpha|\sqrt{\chi_a})$ is introduced, and $g(\chi_a\hat n)=[\hat A,\hat A^{\dagger}]$. The DOCS for this particular case can be obtained as:
      \begin{equation}\label{eq:DOCSMorse}
      |\zeta\rangle =\hat D_f(\alpha)|0\rangle \simeq \sum_{n=0}^{N-1}\left(\begin{array}{c}2N\\n\end{array}\right)^{1/2} \frac{\zeta^n}{(1+|\zeta|^2)^{N}}|n\rangle
      \end{equation}
      where the explicit form of the deformation function has been used and the state is approximate due to the fact that the number of bound states supported by the potential is finite. %This state is then an approximate coherent state in the same sense we found for the AOCS for the Morse potential.

  \section{Numerical results}\label{RES}

  \subsection{Morse Nonlinear coherent states, phase space trajectories and occupation number distribution}
  In this subsection we consider a hetero-nuclear diatomic molecule HF with 22 bound states. Due to the asymmetry of the molecule we model it by means of a Morse potential. According to Carvajal {\em et al} \cite{carvajal}, the position and momentum variables for this system can be expressed as a series expansion involving all powers of the deformed creation and annihilation operators. Keeping up to second-order terms we obtain:
  \begin{equation}\label{eq:xdef}
 \hat x_D\simeq \sqrt{\frac{\hbar}{2\mu\Omega}}\left( f_{00}+f_{10}\hat A^{\dagger}+\hat A f_{01}+f_{20}\hat A^{\dagger 2}+ \hat A^2 f_{02}\right)
  \end{equation}
    \begin{equation}\label{eq:pdef}
 \hat p_D\simeq i\sqrt{\frac{\hbar\mu\Omega}{2}}\left( g_{10}\hat A^{\dagger}+\hat A g_{01}+g_{20}\hat A^{\dagger 2}+ \hat A^2 g_{02}\right)
  \end{equation}
  where the expansion coefficients are functions of the number operator and are given by \cite{reca06,santos13}:
 
  \begin{eqnarray}
  f_{00}(\hat n)&=&  \sqrt{k}\left[f_0+\ln\left(\frac{(k-2)(k-\hat n-1)}{(k-1-2\hat n)(k-2\hat n)}\right)(1-\delta_{\hat n,0})\right] \\
  f_{10}(\hat n)&=&f_{01}(\hat n)=\sqrt{\frac{k-1}{k}}\left(1+\frac{\hat n}{k-\hat n}\right) \\
  f_{20}(\hat n)&=&f_{02}(\hat n)=\frac{k-1}{2k\sqrt{k}}\left(\frac{-1}{(1-(\hat n-1)/k)(1-\hat n/k)}\right), \\
  g_{10}(\hat n)&=&-g_{01}(\hat n)=\sqrt{\frac{k-1}{k}}\left(\frac{k-2\hat n}{k-\hat n}\right),\\
  g_{20}(\hat n)&=&-g_{02}(\hat n)=-\frac{k-1}{k\sqrt{k}}\left(\frac{k-(2\hat n-1)}{k(1-(\hat n-1)/k)(1-\hat n/k)}\right),
  \end{eqnarray}
  where in turn, 
  \begin{equation}
  f_0=\ln k-\left(\sum_{p=1}^{k-2}\frac{1}{p}-\Gamma\right),
  \end{equation}
  with 
  \begin{equation}
  \Gamma = \lim_{m\rightarrow\infty}\left(\sum_{p=1}^{m}\frac{1}{p}-\ln m\right)=0.577216
  \end{equation}
  being the Euler constant and  $k$ is Child's parameter \cite{child} defined by $k=2N+1=1/\chi_a$.
  
  In what follows we define the deformed position and momentum operators $\hat x_D$ and $\hat p_D$  taking $\hbar=\mu=\Omega=1$ in Eqs.~\ref{eq:xdef} and \ref{eq:pdef}. When the deformation function is equal to 1 ($\chi_a=0$), these expressions take the harmonic values. In Ref.~\cite{reca06} we compared the phase space trajectories obtained by averaging the deformed coordinate $\hat x_D$ and momentum $\hat p_D$ (keeping up to second order terms) with those obtained averaging the Morse coordinate and momentum and we found a very good agreement between them when the averages were taken between nonlinear coherent states obtained as eigenstates of the deformed annihilation operator (AOCS).
%  \subsection{Phase space trajectories and occupation number distribution}
 In order to evaluate the phase space trajectories, here we will consider the temporal evolution of the nonlinear coherent states obtained by application of the deformed displacement operator on the vacuum state (see Eq.~\ref{eq:DOCSMorse}). That is, we apply the time evolution operator $\hat U(t)=e^{-i\hat H_D t/\hbar}$ on the state $|\zeta\rangle$.
 \begin{equation}
 |\zeta;t\rangle = \hat U(t)|\zeta\rangle = e^{-i\Omega t\left[\hat n+\frac{1}{2}-\chi_a\left(\hat n+\frac{1}{2}\right)^2-\frac{\chi_a}{4}\right]}|\zeta\rangle.
 \end{equation}
The averages are:
\begin{equation}
\langle \zeta,t|\hat x_D|\zeta,t\rangle = \langle \zeta|\hat U(t)^{\dagger}\hat x_D \hat U(t)|\zeta\rangle, \ \ \langle \zeta,t|\hat p_D|\zeta,t\rangle = \langle \zeta|\hat U(t)^{\dagger}\hat x_D \hat U(t)|\zeta\rangle.
\end{equation}
Transformation of the deformed operators yield:
\begin{eqnarray}
\hat U(t)^{\dagger}\hat A \hat U(t)&=&  e^{i\Omega t\left[\hat n-\chi_a\left(\hat n^2+\hat n\right)\right]}\hat A  e^{-i\Omega t\left[\hat n-\chi_a\left(\hat n^2+\hat n\right)\right]} \nonumber \\ &=& e^{-i\Omega\chi_a t\hat n^2}e^{i\Omega t(1-\chi_a)\hat n}\hat A e^{-i\Omega t(1-\chi_a)\hat n}e^{i\Omega\chi_a t\hat n^2}\nonumber \\ &=& e^{-i\Omega (1-\chi_a)t}e^{-i\Omega\chi_a t\hat n^2}\hat A e^{i\Omega\chi_a t\hat n^2} = e^{-i\Omega t (1-2\chi_a)}e^{-2i\Omega\chi_a t \hat n}\hat A
\end{eqnarray}    
and
\begin{equation}
\hat U(t)^{\dagger}\hat A^{\dagger} \hat U(t)=e^{i\Omega t (1-2\chi_a)}\hat A^{\dagger} e^{2i\Omega\chi_a t \hat n},
\end{equation}
notice the presence of the number operator in the exponentials. From these expressions we can get $\hat A^2(t)=\hat U(t)^{\dagger} \hat A^2 \hat U(t)$ and $\hat A^{\dagger 2}(t)=\hat U(t)^{\dagger}\hat A^{\dagger 2}\hat U(t)$ and obtain the deformed coordinate and momentum as a function of time. 

\begin{figure}[htb]
\begin{center}
\includegraphics[width=7cm, height=5cm]{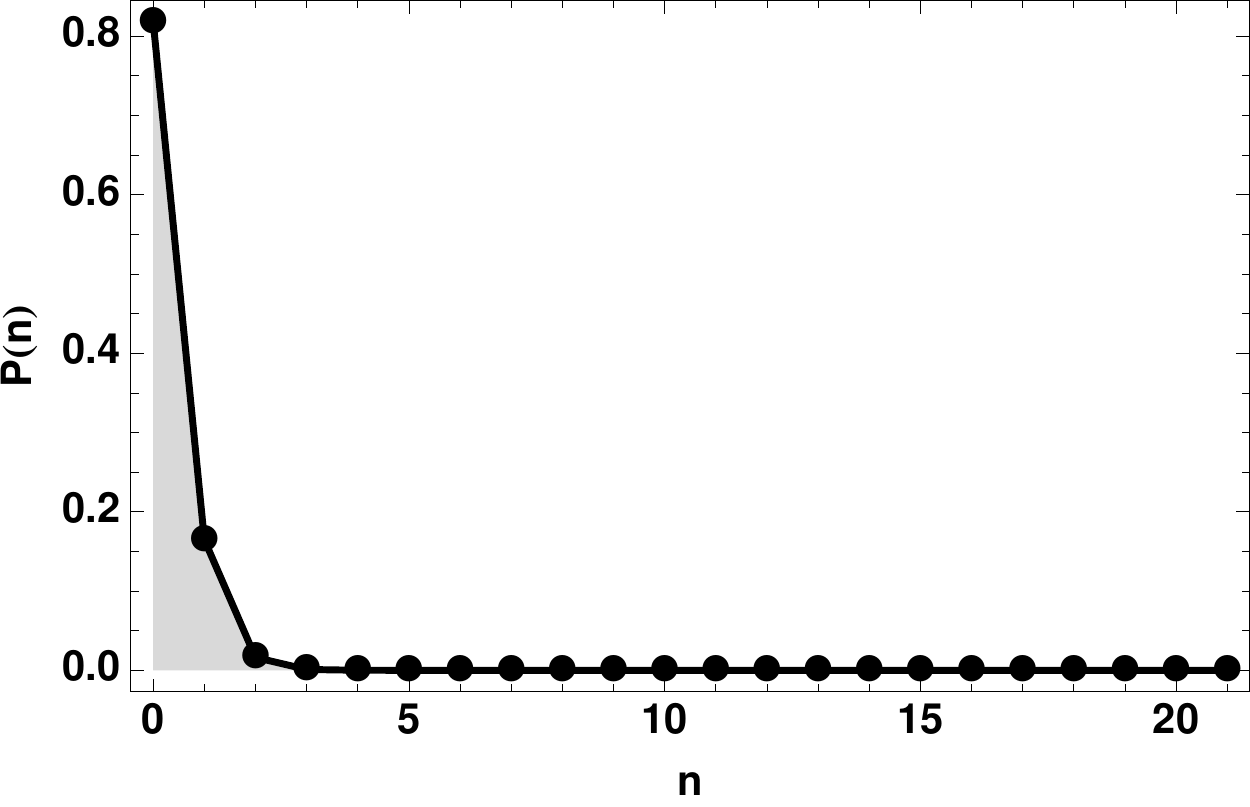} 
\includegraphics[width=7cm, height=6.5cm]{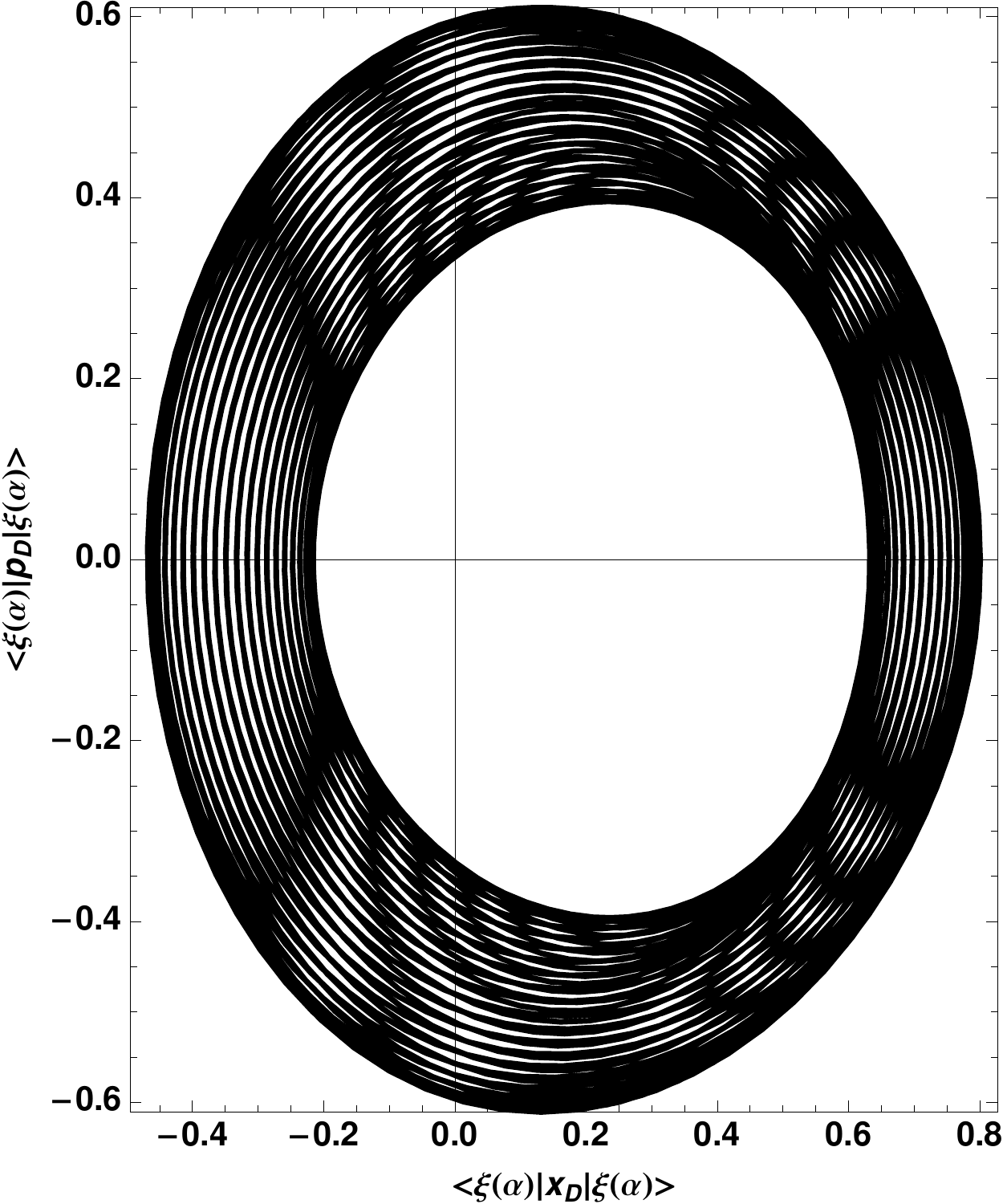} 
\includegraphics[width=7cm, height=5cm]{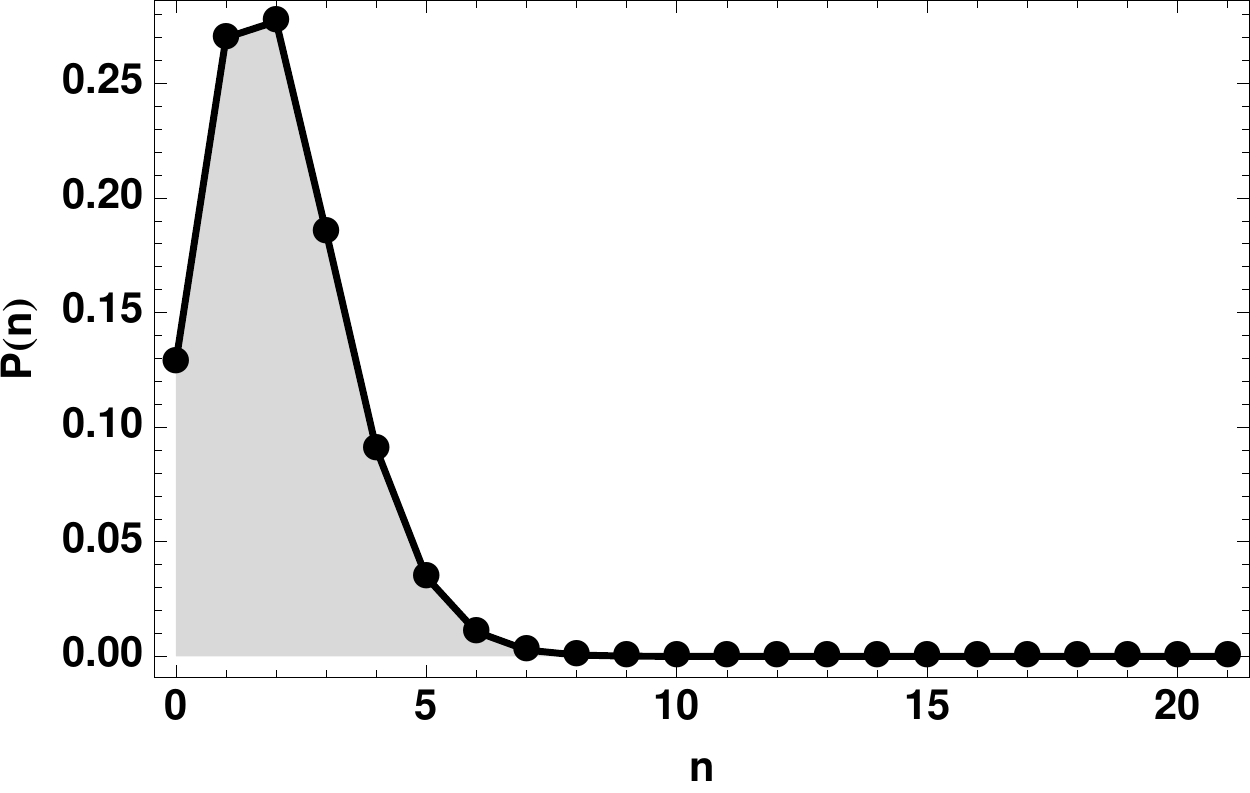} 
\includegraphics[width=7cm, height=6.5cm]{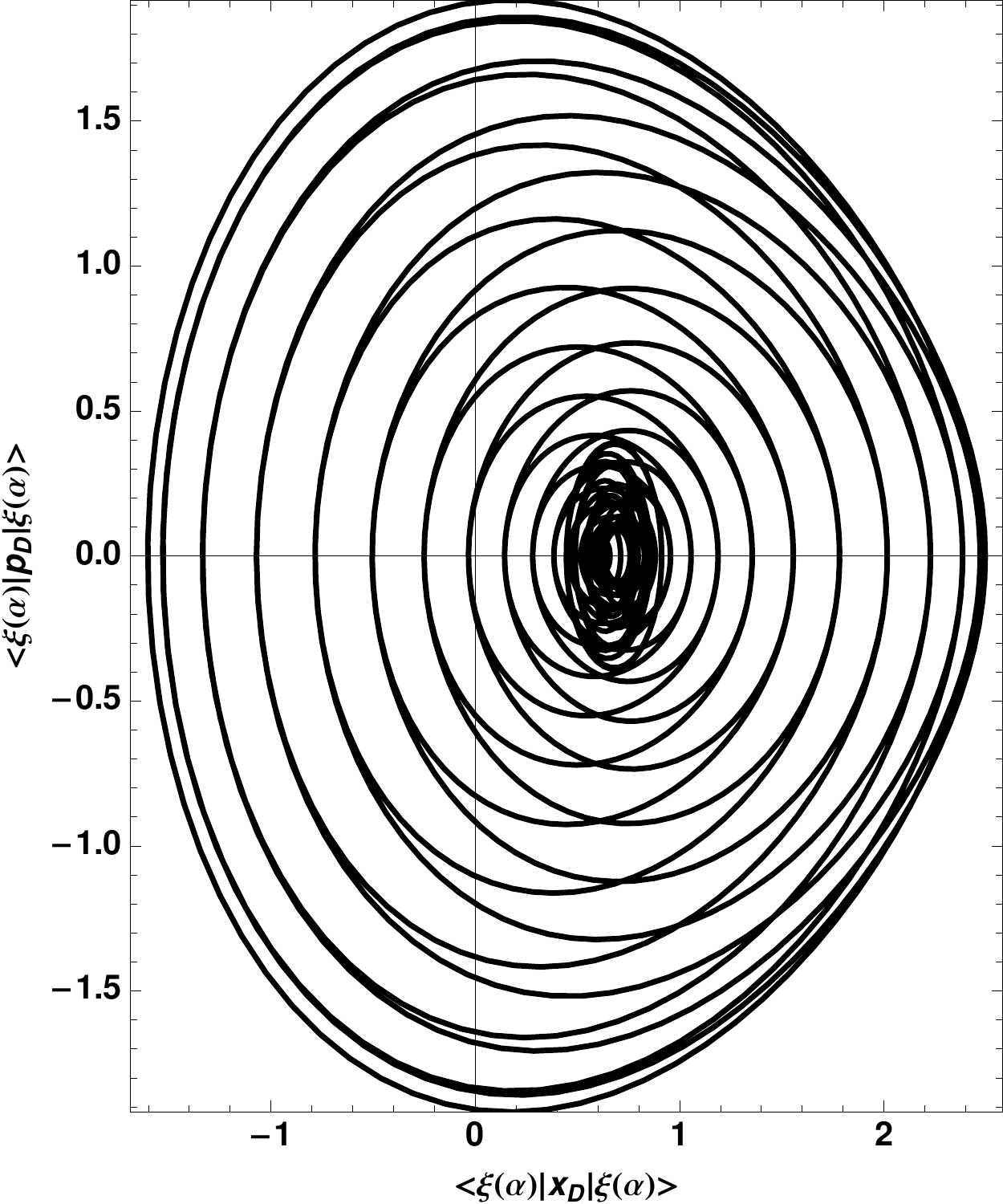} 
\includegraphics[width=7cm, height=5cm]{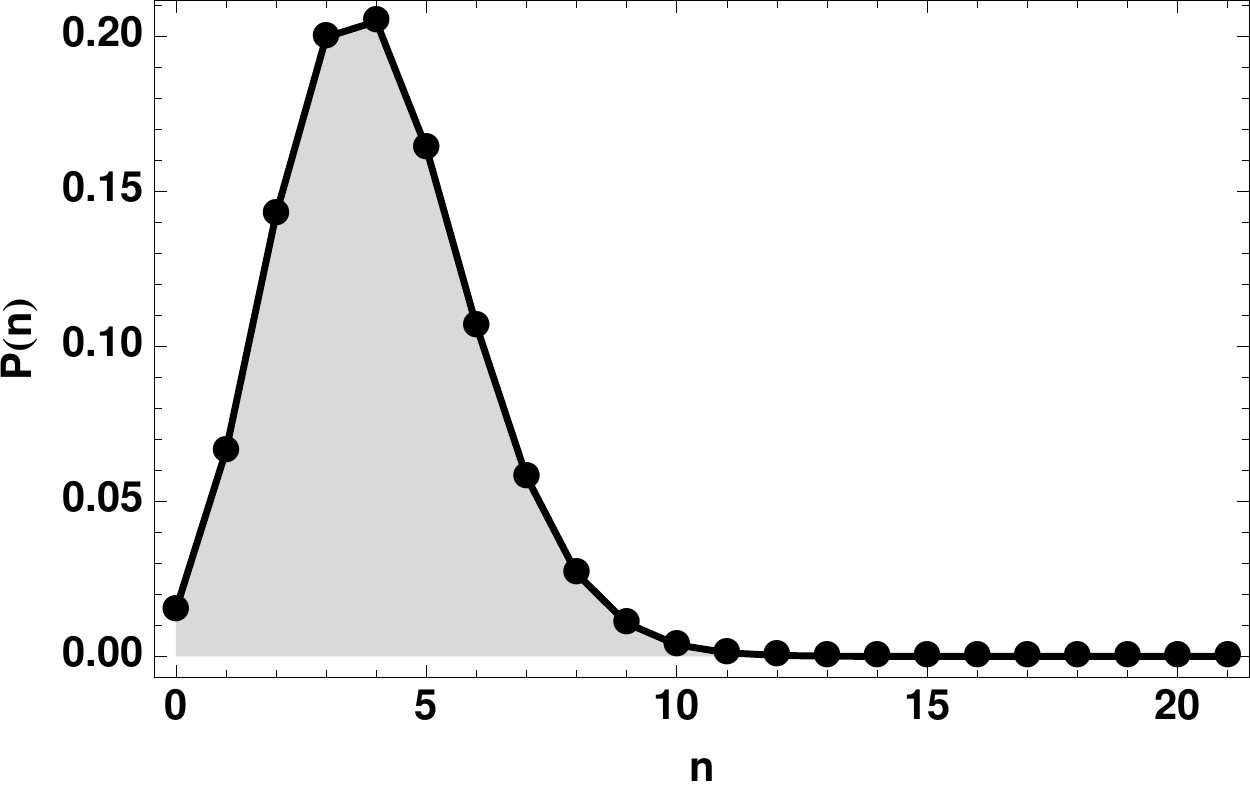} 
\includegraphics[width=7cm, height=6.5cm]{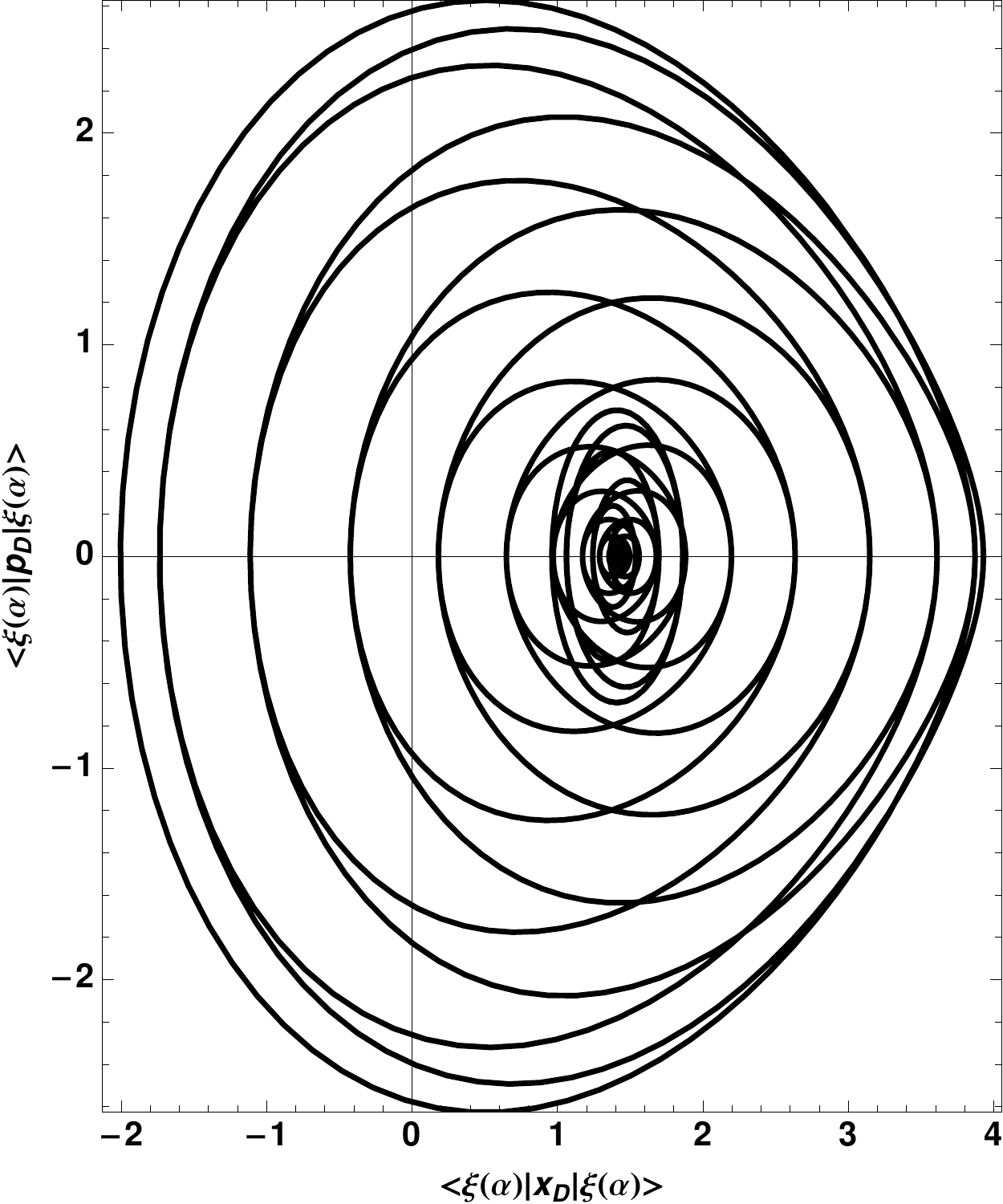} 
\end{center}
\caption{Occupation number distributions (left column) and the corresponding phase space trajectories (right column) of deformed displacement operator coherent states (DOCS) for $\langle \hat{n} \rangle = 0.2$, $2$, and $4$. In the calculation, a diatomic HF molecule is considered to be modeled by a deformed Morse-like oscillator with $N=22$ bound states.}
\label{fig:phasespace}
\end{figure}

In figure \ref{fig:phasespace} we show the occupation number distributions $P_n(\alpha)=|\langle n|\zeta(\alpha)\rangle|^2$ (left column) and the corresponding phase space trajectories (right column) for $\langle \hat{n} \rangle = 0.2$, $2$, and $4$. In the calculation, a diatomic HF molecule is considered to be modeled by a deformed Morse-like oscillator with $N=22$ bound states. It can be seen from the figure that for the values used for the parameter $\alpha$, which fixes the average value of the number operator, the occupation number distributions are such that the states near dissociation are mostly unoccupied so it is to be expected that the contribution from the states in the continuum can be neglected. Concerning the phase space trajectories we see that for a given value of the average number operator, that is, a given value of the average energy, there are  several intersecting curves in contrast with the case of a field coherent state where one finds a single trajectory for a given energy. The presence of several trajectories for a given energy is a signature of the nonlinear term in the deformed Hamiltonian. Notice that for a small energy (right column top) the phase space trajectories fill an anular region of phase space with an unaccessible internal region. The width of the anular region narrows when the parameter $|\alpha|$ is decreased, that is, for smaller values of the average number operator. This unaccessible region is lost as the energy is increased (right column medium and bottom). Due to the asymmetry of the Morse potential the deformed coordinate can attain small negative values and much larger positive values along the temporal evolution.

\begin{figure}[htb]
\begin{center}
\includegraphics[width=7cm, height=5cm]{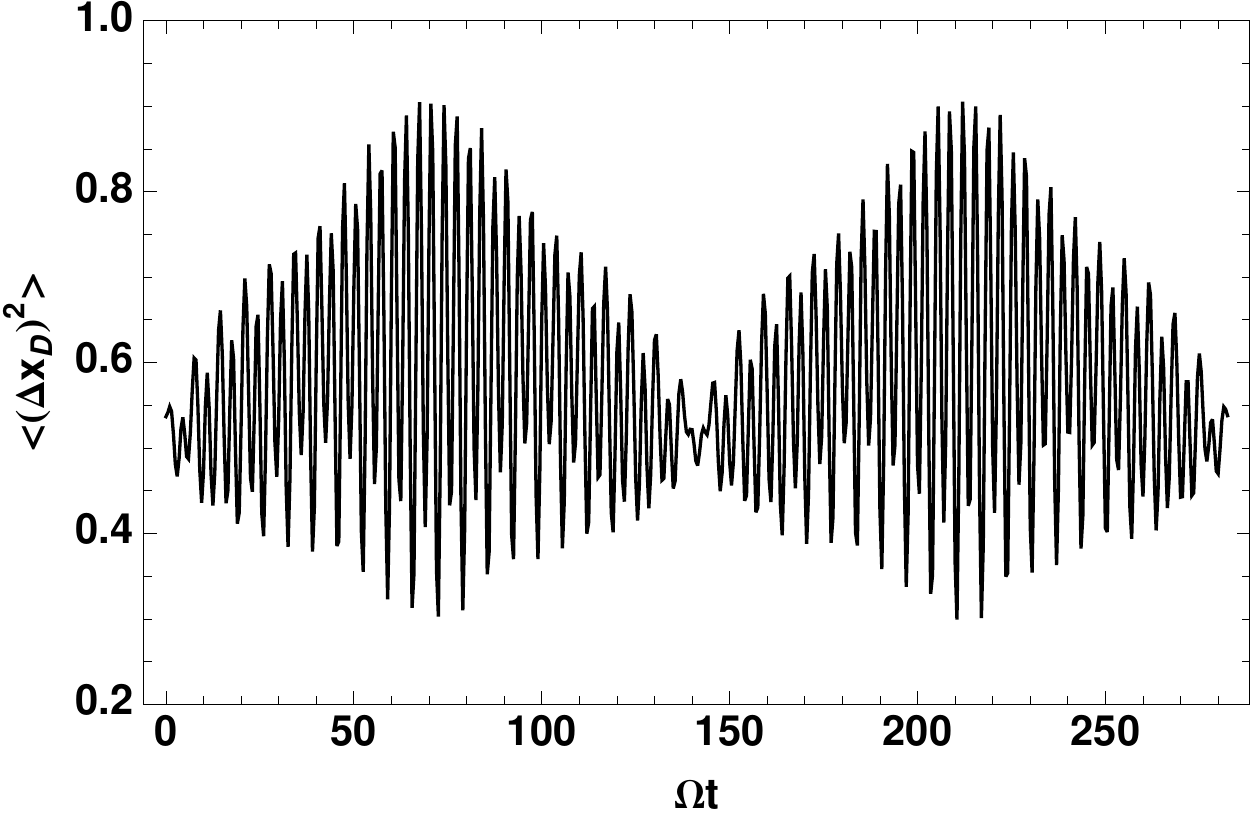} 
\includegraphics[width=7cm, height=5cm]{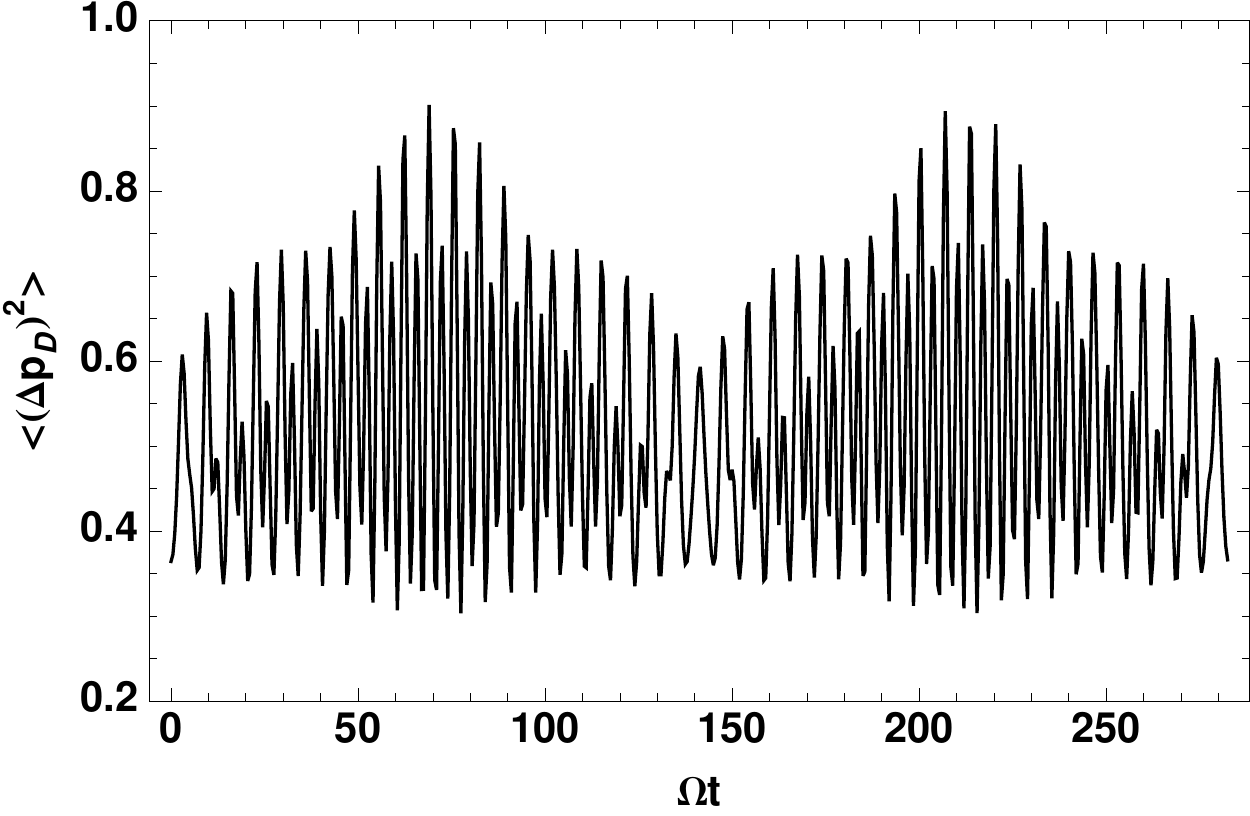} 
\includegraphics[width=7cm, height=5cm]{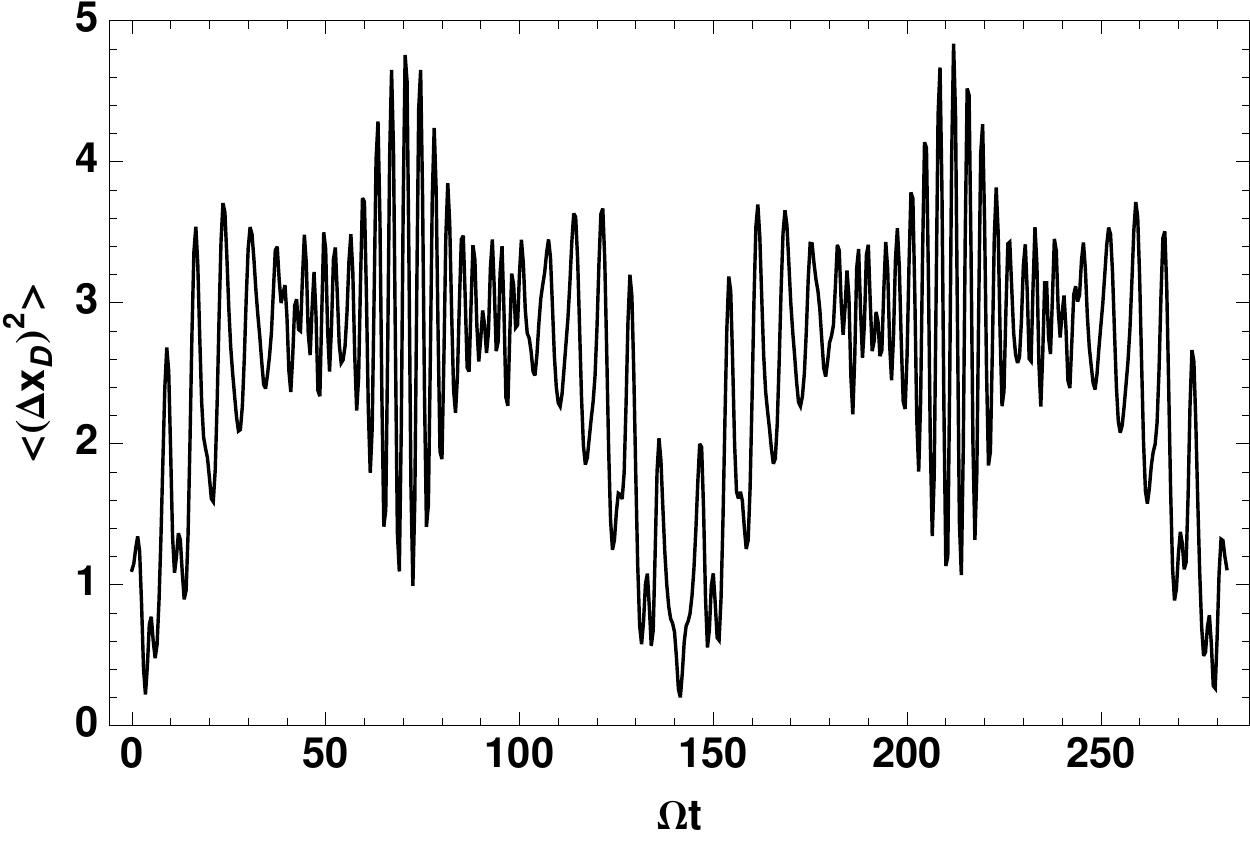} 
\includegraphics[width=7cm, height=5cm]{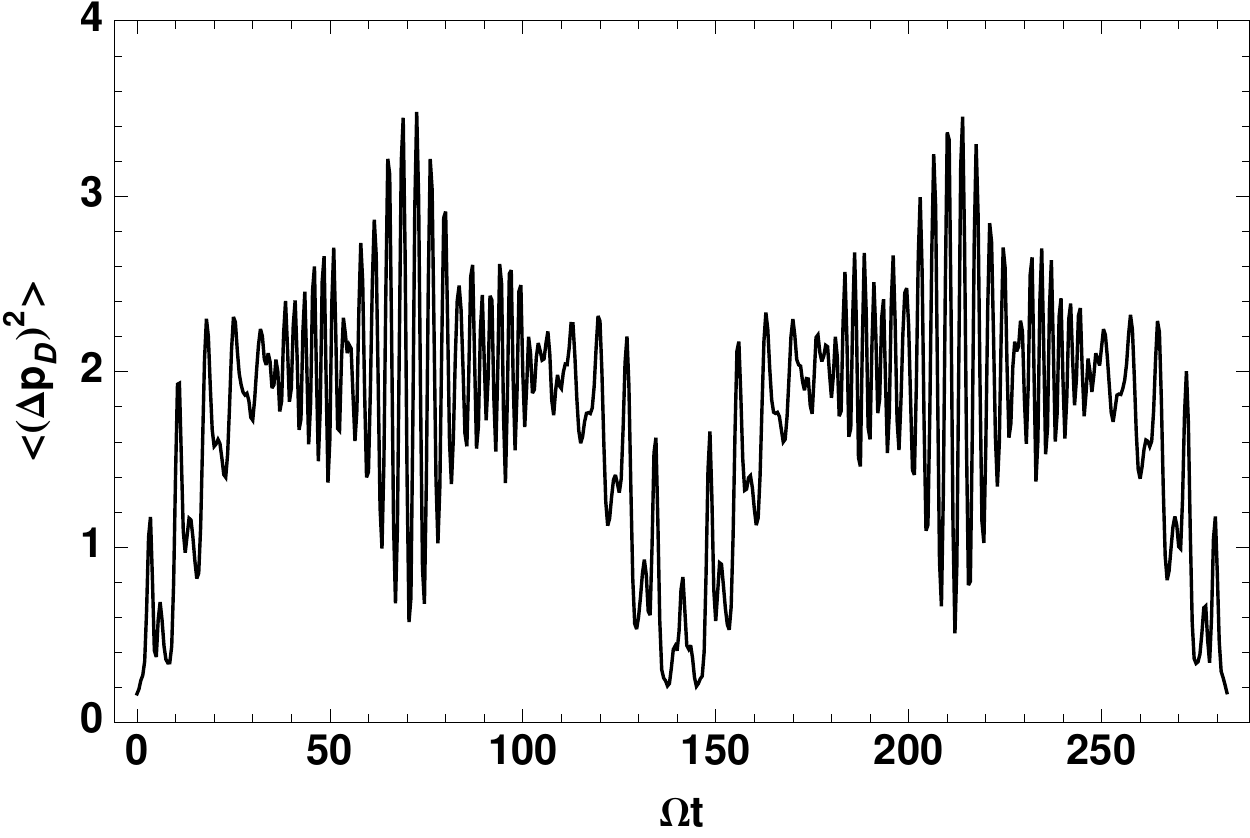} 
\includegraphics[width=7cm, height=5cm]{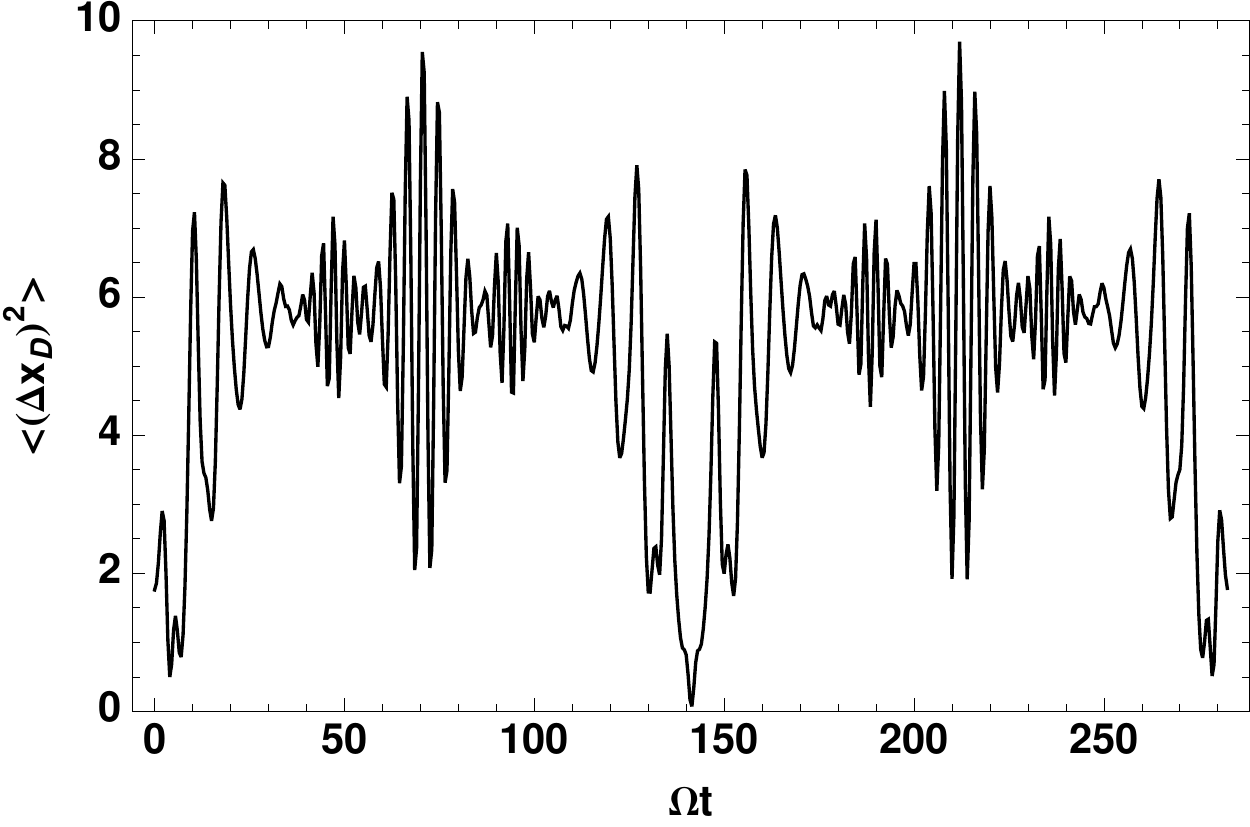} 
\includegraphics[width=7cm, height=5cm]{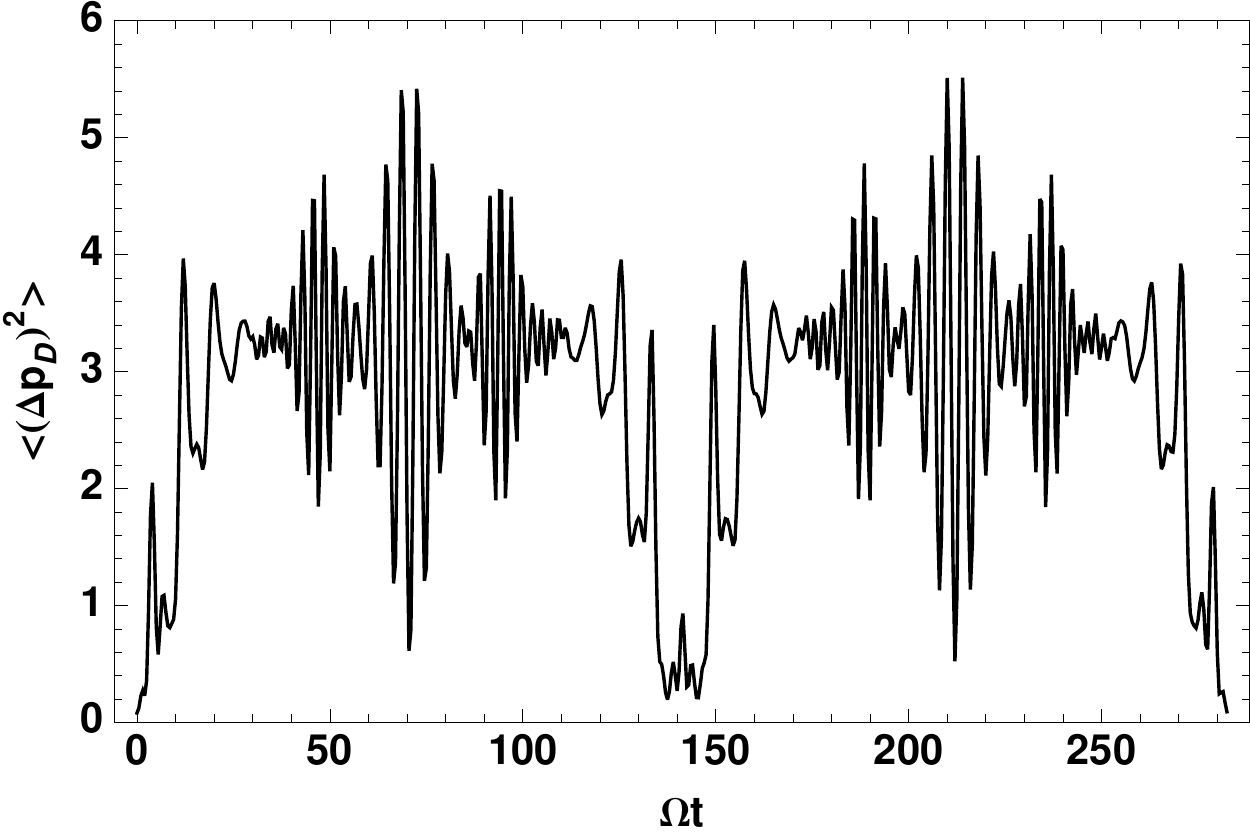} 
\end{center}
\caption{Temporal evolution of the uncertainties in coordinate (left column) and momentum (right column) of deformed displacement operator coherent states (DOCS) for $\langle \hat{n} \rangle = 0.2$, $2$, and $4$. In the calculation, a diatomic HF molecule is considered to be modeled by a deformed Morse-like oscillator with $N=22$ bound states.}
\label{fig:uncertainties}
\end{figure}

In figure \ref{fig:uncertainties} we show the temporal evolution of the uncertainties in coordinate (left column) and momentum (right column) of deformed displacement operator coherent states (DOCS) for $\langle \hat{n} \rangle = 0.2$, $2$, and $4$.  The first row corresponds to the case with lowest energy considered $\langle \hat n\rangle=0.2$, the dispersions in the deformed position and momentum are oscillating functions with amplitudes in the range $0.3 \leq \langle (\Delta \hat O)^2 \rangle \leq 0.9$ with $\hat O = \hat x_D, \hat p_D$. Notice that the DOCS are not minimum uncertainty states and notice also that there is squeezing present whenever the dispersion in any coordinate is smaller than 0.5. The second  row corresponds to $\langle \hat n\rangle=2$, here the dispersions  oscillate with larger amplitude  $0.2 \leq(\Delta \hat O)^2\leq 4.8$. And finally, in the third row, $\langle \hat n\rangle=4$, the dispersions  oscillate with amplitude in the range $0.1 \leq(\Delta \hat O)^2\leq 10$. In the cases with $\langle \hat n\rangle=2, 4$ the amplitude of the oscillations in the dispersion in the coordinate is larger than that in the momentum.
  
\begin{figure}[htb]
\begin{center}
\includegraphics[width=11cm, height=15cm]{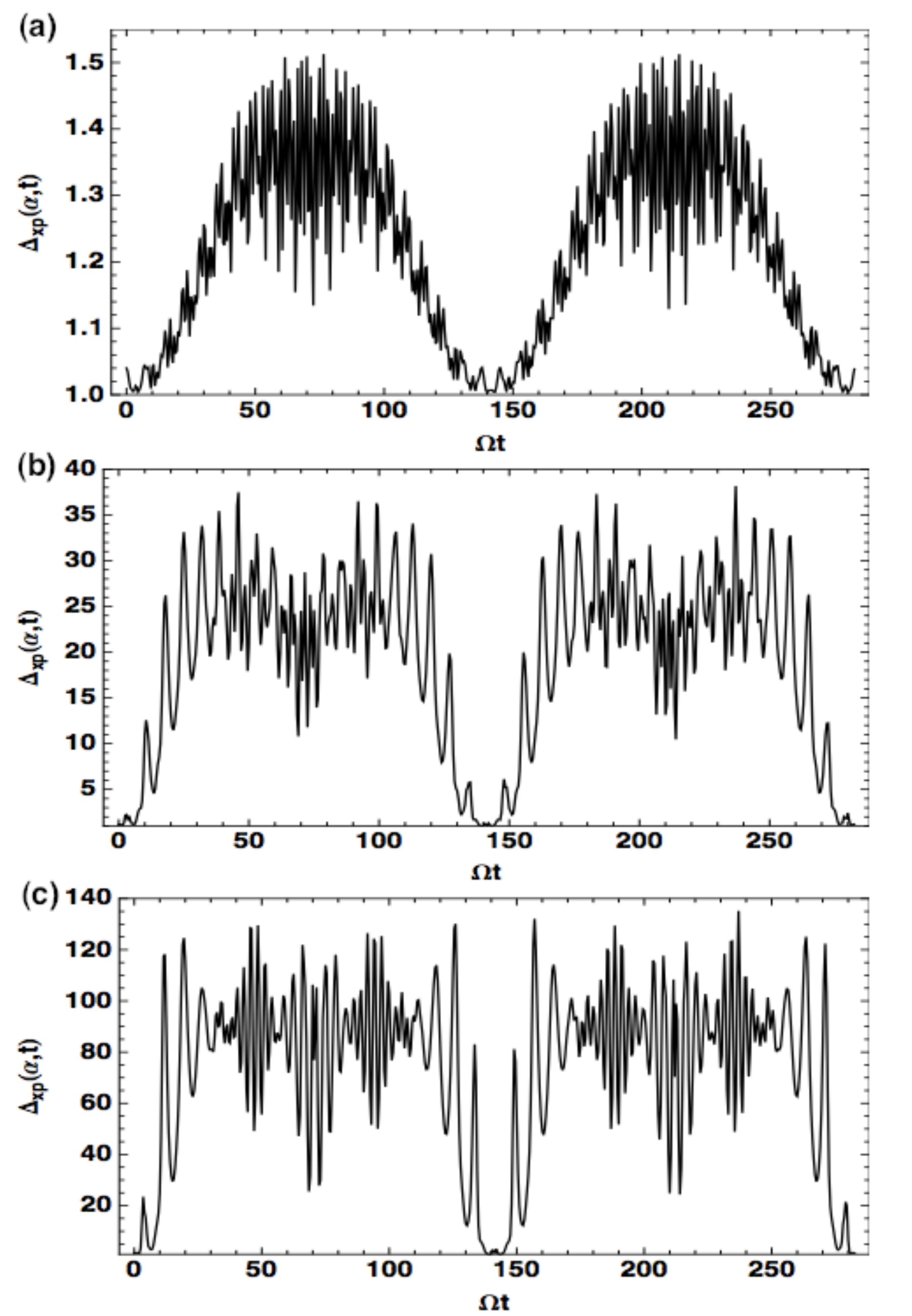} 
\end{center}
\caption{Temporal evolution of the normalized uncertainty product $\Delta_{xp} = 4\langle (\Delta \hat x_{D})^{2} \rangle \langle (\Delta \hat p_{D})^{2} \rangle/|\langle [\hat x_{D},\hat p_{D}] \rangle|^{2}$ of displacement operator coherent states for $\langle \hat{n} \rangle = 0.2$, $2$, and $4$ (frames (a), (b) and (c), respectively). In the calculation, a diatomic HF molecule is considered to be modeled by a deformed Morse-like oscillator with $N=22$ bound states.}
\label{fig:puncertainties}
\end{figure}

  In figure \ref{fig:puncertainties} we show the temporal evolution of the normalized uncertainty product $\Delta_{xp} = 4\langle (\Delta \hat x_{D})^{2} \rangle \langle (\Delta \hat p_{D})^{2} \rangle/|\langle [\hat x_{D},\hat p_{D}] \rangle|^{2}$ of displacement operator coherent states for $\langle \hat{n} \rangle = 0.2$, $2$, and $4$ (frames (a), (b) and (c), respectively). We see that there are some specific times at which the dispersion is minimal and these times do not depend on the value of $\langle \hat{n} \rangle$, these correspond to the outermost trajectories shown in figure \ref{fig:phasespace}. At those times when the trajectories evolve near the origin, so that $\langle \hat x_D\rangle$ and $\langle \hat p_D\rangle$ are small, the dispersions take their largest values \cite{reca06}. Most of the time the DOCS are not minimum uncertainty states, the product of the dispersions seems to be a periodic function of time and the amount of dispersion is an increasing function of $\langle \hat{n} \rangle$.
  
\begin{figure}[htb]
\begin{center}
\includegraphics[width=11cm, height=15cm]{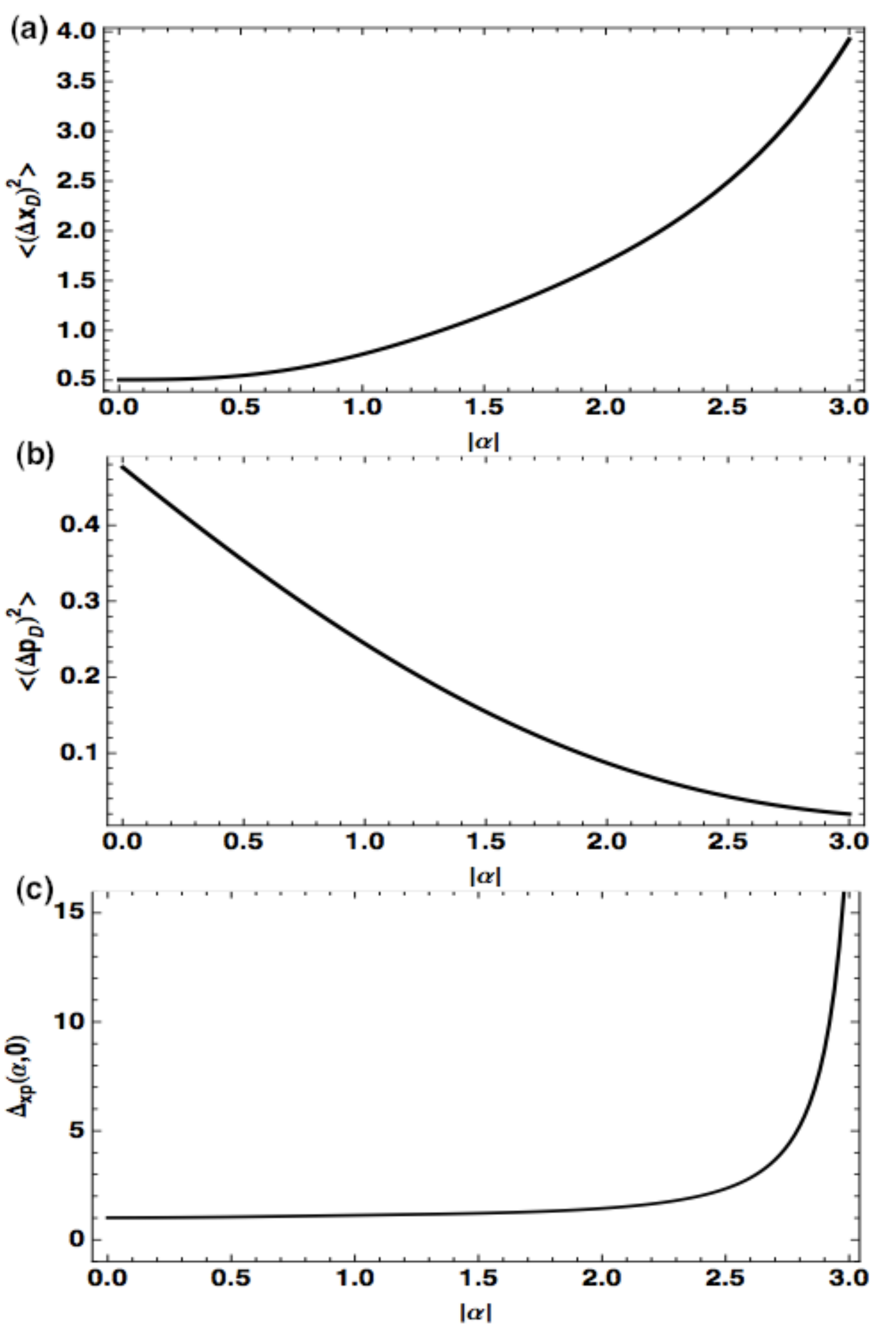} 
\end{center}
\caption{Uncertainty in coordinate $\langle (\Delta \hat x_{D})^{2} \rangle$ (a), in momentum $\langle (\Delta \hat p_{D})^{2} \rangle$ (b), and the corresponding normalized uncertainty product $\Delta_{xp} = 4\langle (\Delta \hat x_{D})^{2} \rangle \langle (\Delta \hat p_{D})^{2} \rangle/|\langle [\hat x_{D},\hat p_{D}] \rangle|^{2}$ (c) at time $t=0$ as functions of $|\alpha|$ for the DOCS. In the calculation, a diatomic HF molecule is considered to be modeled by a deformed Morse-like oscillator with $N=22$ bound states.}
\label{fig:uncertaintiesalpha}
\end{figure}

In figure \ref{fig:uncertaintiesalpha} we show the dispersions in coordinate and momentum at time $t=0$ as a function of the parameter $|\alpha|$. Here we see that the dispersion in the deformed coordinate $\hat x_D$ starts as that of a minimum uncertainty state and is an increasing function of the parameter $|\alpha|$. The dispersion in the momentum also starts as that of a minimum uncertainty state and is a decreasing function of the parameter $|\alpha|$ so it is squeezed. The normalized uncertainty product remains near a minimum uncertainty state for small values of $|\alpha|$. However, for values of $|\alpha|$ such that the average value of the number operator approaches $N$ the uncertainty product increases rapidly. 

\subsection{Modified P\"oschl-Teller Nonlinear coherent states, phase space trajectories and occupation number distribution}
%\subsection{Modified P\"oschl-Teller Non linear coherent states}
In this subsection we consider a homonuclear diatomic molecule $H_2$ supporting 10 bound states. Due to the symmetry of the potential, the deformed coordinate and momentum are written as an expansion involving odd powers   of deformed operators \cite{lemus} as:
\begin{equation}
\hat X_D=\sqrt{\frac{\hbar}{2\mu\omega}}\left(\hat A F(\hat n)+ F(\hat n)\hat A^{\dagger}+\hat A^3 G(\hat n)+ G(\hat n)\hat A^{\dagger 3}\right)
\end{equation}
\begin{equation}
\hat P_D=-i \sqrt{\frac{\hbar\mu\omega}{2}}\left(\hat A R(\hat n)- R(\hat n)\hat A^{\dagger}+\hat A^3 S(\hat n)- S(\hat n)\hat A^{\dagger 3}\right)
\end{equation}
where we have kept up to third order terms in the deformed operators and the coefficient functions are given by:
\begin{equation}
F(n)=\sqrt{\frac{2\mu\omega}{\hbar}}\frac{\langle n-1|\hat x|n\rangle}{f(n)\sqrt{n}},
\end{equation}
\begin{equation}
G(n)=\sqrt{\frac{2\mu\omega}{\hbar}}\frac{\langle n-3|\hat x|n\rangle}{f(n)f(n-1)f(n-2)\sqrt{n(n-1)(n-2)}},
\end{equation}
\begin{equation}
R(n)= i\sqrt{\frac{2}{\hbar\mu\omega}}\frac{\langle n-1|\hat p|n\rangle}{f(n)\sqrt{n}},
\end{equation}
\begin{equation}
S(n)=i\sqrt{\frac{2}{\hbar\mu\omega}}\frac{\langle n-3|\hat p|n\rangle}{f(n)f(n-1)f(n-2)\sqrt{n(n-1)(n-2)}}.
\end{equation}
Here, the matrix elements $\langle n-\beta|\hat x|n\rangle$ and $\langle n-\beta|\hat p|n\rangle$ are evaluated by numerical integration using the eigenfunctions of the corresponding Scr\"odinger equation.

The temporal evolution of the deformed coordinate and momentum is calculated taking the expectation values between the states $|\zeta,t\rangle=\hat U(t)|\zeta\rangle$ with $|\zeta\rangle$ a nonlinear coherent state obtained by application of the deformed displacement operator on the vacuum state
\begin{equation}
|\zeta\rangle \simeq \frac{1}{(1+|\zeta|^2)^s}\sum_{n=0}^{s-1} \sqrt{\frac{\Gamma(2s+1)}{n!\Gamma(2s+1-n)}} \zeta^{n}|n\rangle,
\end{equation}
(the state is approximate since the sum contains a finite number of terms) and
\[ \hat U(t)=\exp(-\frac{i}{\hbar}\hat H_D t)\] with
\[ \hat H_D=\frac{\hbar^2 a^2}{2\mu}(-\hat n^2+2s\hat n+s).\]

\begin{figure}[htb]
\begin{center}
\includegraphics[width=7cm, height=5.5cm]{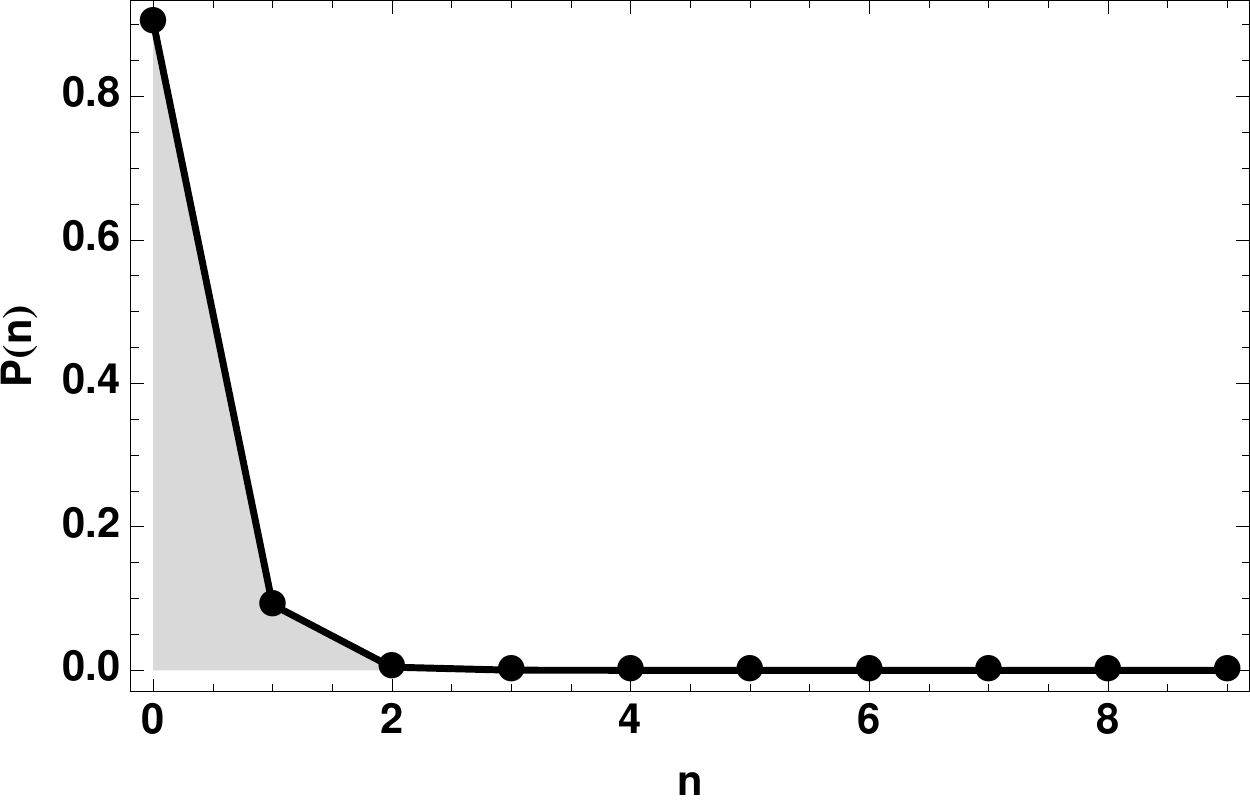} 
\includegraphics[width=5.5cm, height=7.5cm]{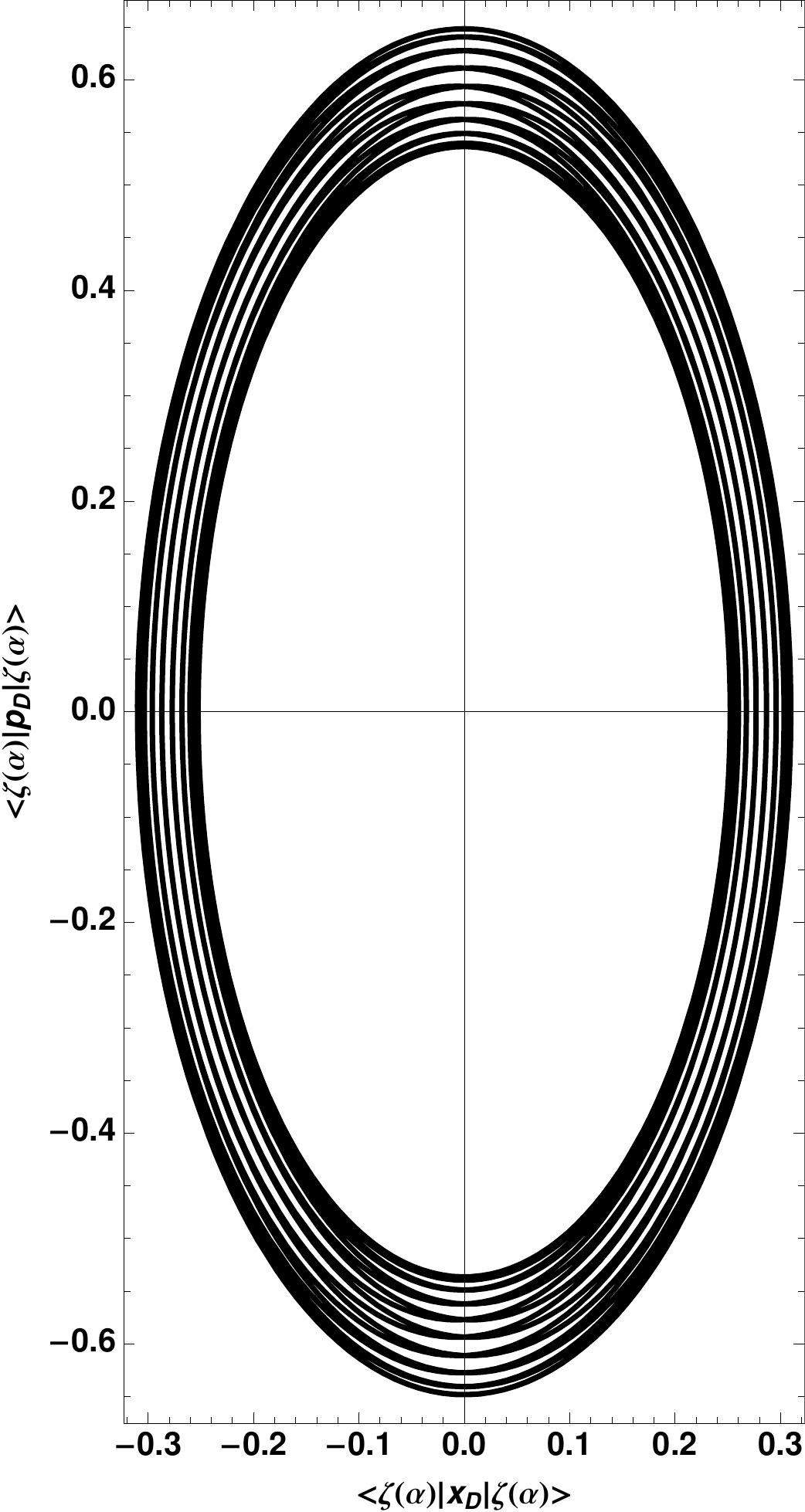} 
\includegraphics[width=7cm, height=5.5cm]{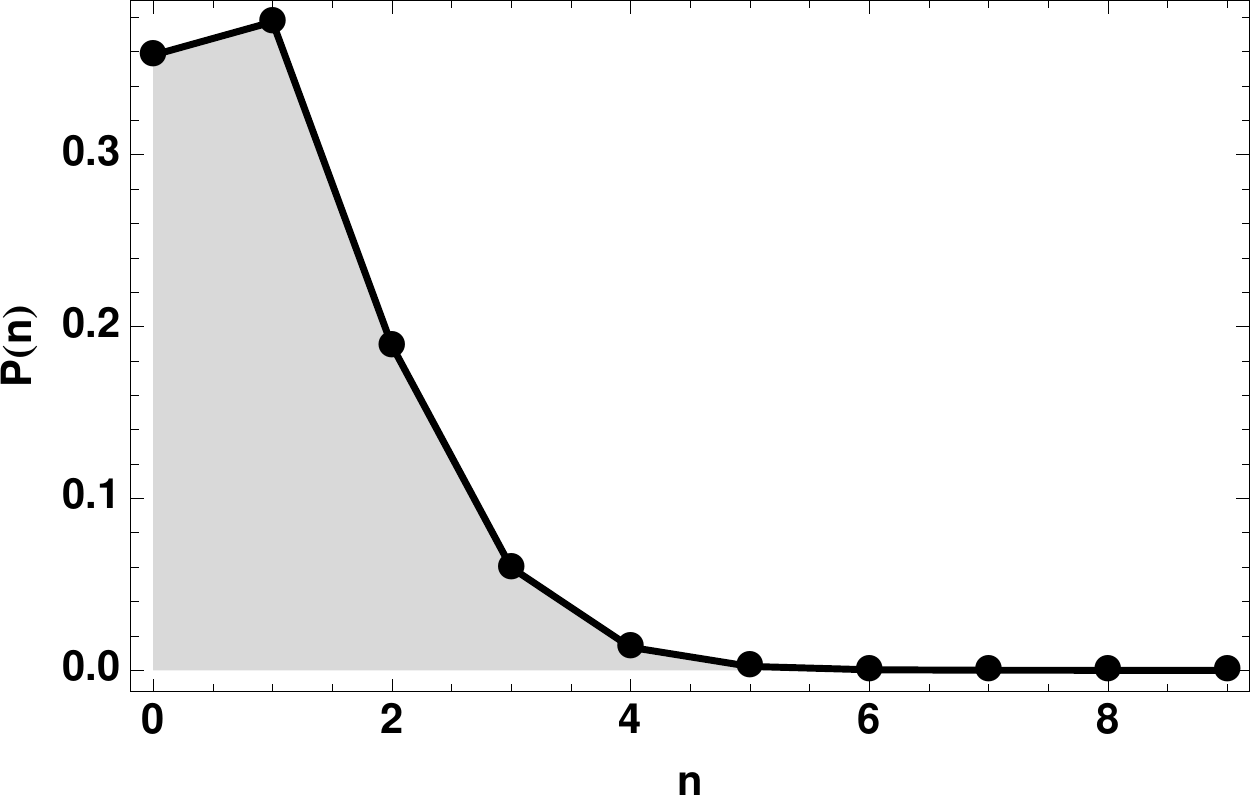} 
\includegraphics[width=5.5cm, height=7.5cm]{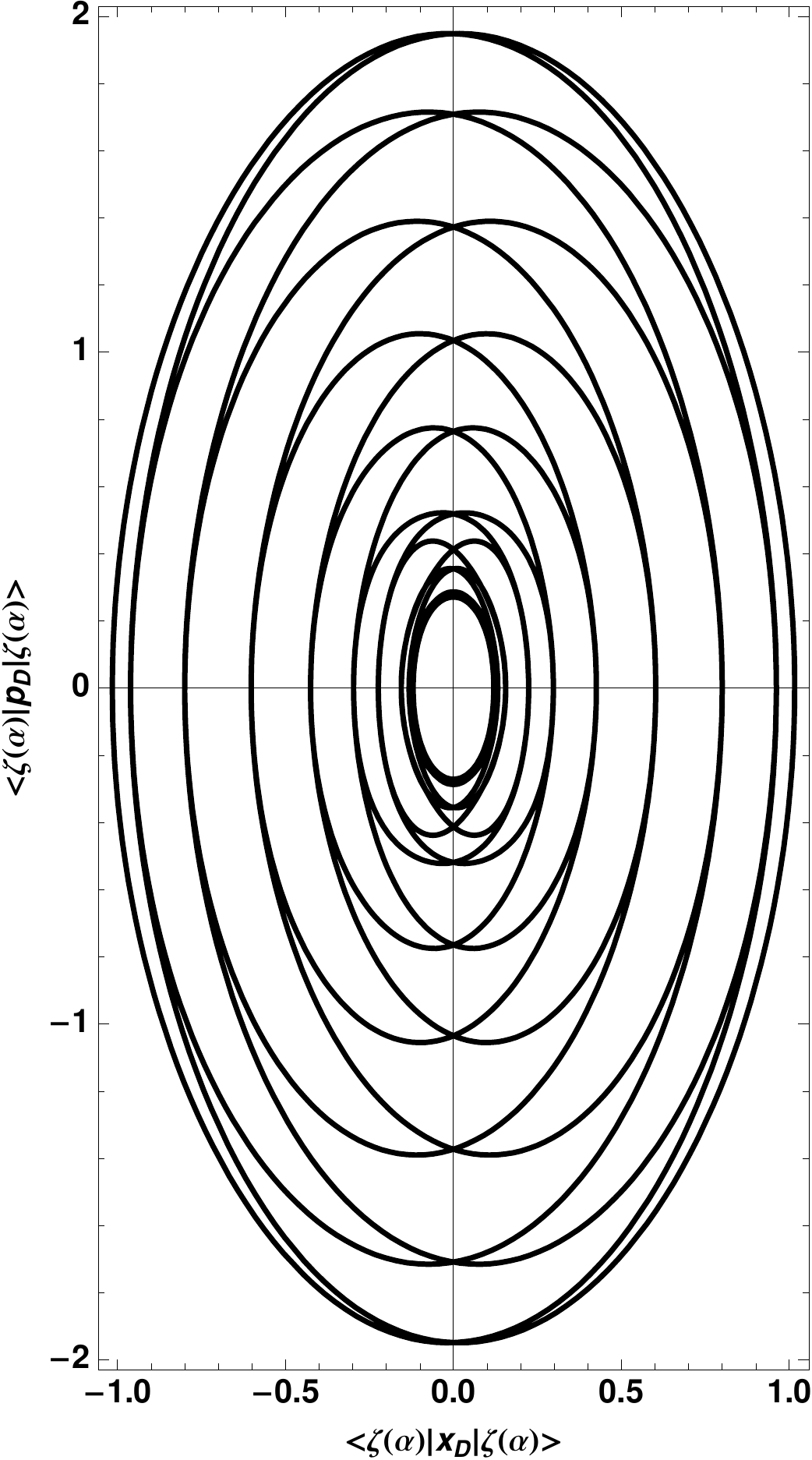} 
\end{center}
\caption{Occupation number distributions (left column) and the corresponding phase space trajectories (right column) of deformed displacement operator coherent states (DOCS) for $\langle \hat{n} \rangle \approx 0.1$, $1.0$. In the calculation, a modified P\"oschl-Teller potential is modeled by a deformed oscillator with $s=10$ bound states.}
\label{fig:pspace}
\end{figure}

In figure \ref{fig:pspace} we show the occupation number distributions and the phase space trajectories of DOCS for the particular case of $\langle \hat n\rangle\simeq 0.1$ and $1.0$. Notice that the phase space trajectories are symmetric with respect to the origin, this is a reflection of the symmetry of the potential. For a small value of the parameter $|\alpha|$ we see a behavior similar to that found for the Morse potential (see figure \ref{fig:phasespace}), that is, the phase space trajectories fill an anular region whose width depends upon the parameter $|\alpha|$ (for smaller values of $|\alpha|$ the width decreases) and  there is an unaccessible internal region. For a larger value of the parameter $|\alpha|$ the trajectories in phase space are intersecting curves occupying all the space consistent with the energy, the unaccessible region disappears. The plots of the occupation number distribution manifest the fact that for the cases considered here the population is far from the dissociation and the approximation done keeping only the bounded part of the spectrum is justified. 

\begin{figure}[htb]
\begin{center}
\includegraphics[width=14cm, height=15cm]{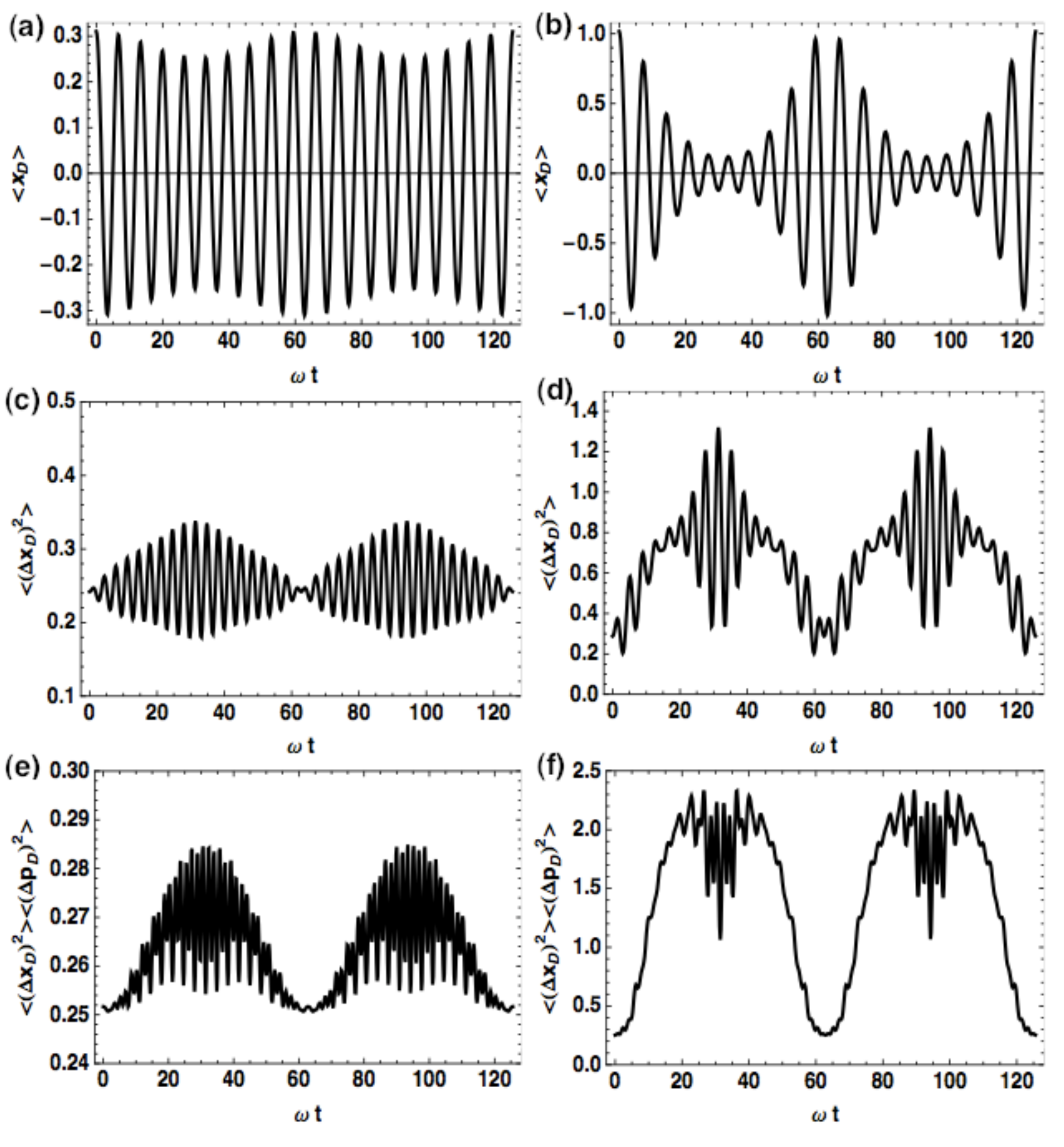} 
\end{center}
\caption{Average coordinate (frames (a) and (b)), uncertainty in coordinate (frames (c) and (d)), and the corresponding uncertainty product (frames (e) and (f)) of deformed displacement operator coherent states (DOCS) for $\langle \hat{n} \rangle \approx 0.1$ (left column) and $1.0$ (right column). In the calculation, a modified P\"oschl-Teller potential is modeled by a deformed oscillator with $s=10$ bound states.}
\label{fig:uncertainties-rm}
\end{figure}

In figure \ref{fig:uncertainties-rm} we show the temporal evolution of the average value of the deformed coordinate $\langle \hat X_D\rangle$, its dispersion $\langle (\Delta \hat X_D)^2\rangle$ and the uncertainty product $\langle (\Delta \hat X_D)^2\rangle \langle (\Delta \hat P_D)^2\rangle$ for $\langle \hat n\rangle\simeq 0.1$ (left column) and $1.0$ (right column). For these calculations we considered a system supporting 10 bound states.
For a fixed, small value of the parameter $|\alpha|$ (thus a small $\langle \hat n\rangle$) the deformed coordinate is an oscillatory function with a slightly varying amplitude, the corresponding dispersion is an oscillatory function whose amplitude is largest when the amplitude of the oscillation is smallest. Notice the presence of squeezing. The corresponding uncertainty product is shown in frame (e), it can be seen that the DOCS we have constructed are minimum uncertainty states at the initial time and there are some specific instants of time when the states return to be of minimum uncertainty, this conduct is periodic. Most of the time the DOCS are not minimum uncertainty states.\\
For a larger value of the parameter $|\alpha|$ (thus a larger value of the average $\langle \hat n\rangle$) we see that the deformed coordinate presents oscillations with varying amplitudes. The largest amplitude corresponds to that of a field coherent state with the same energy. Notice that when the amplitude is largest the dispersion is smallest and corresponds to that of a minimum uncertainty state, this conduct is present at the initial time and repeats itself periodically. Here we can see also the presence of squeezing. When the amplitude of the oscillations in the deformed coordinate is small we see that the dispersion is large, this is a reflection of the nonlinear terms in the Hamiltonian. The uncertainty product also shows a periodic conduct with instants of time where the state is a minimum uncertainty state.  

\begin{figure}[htb]
\begin{center}
\includegraphics[width=11cm, height=15cm]{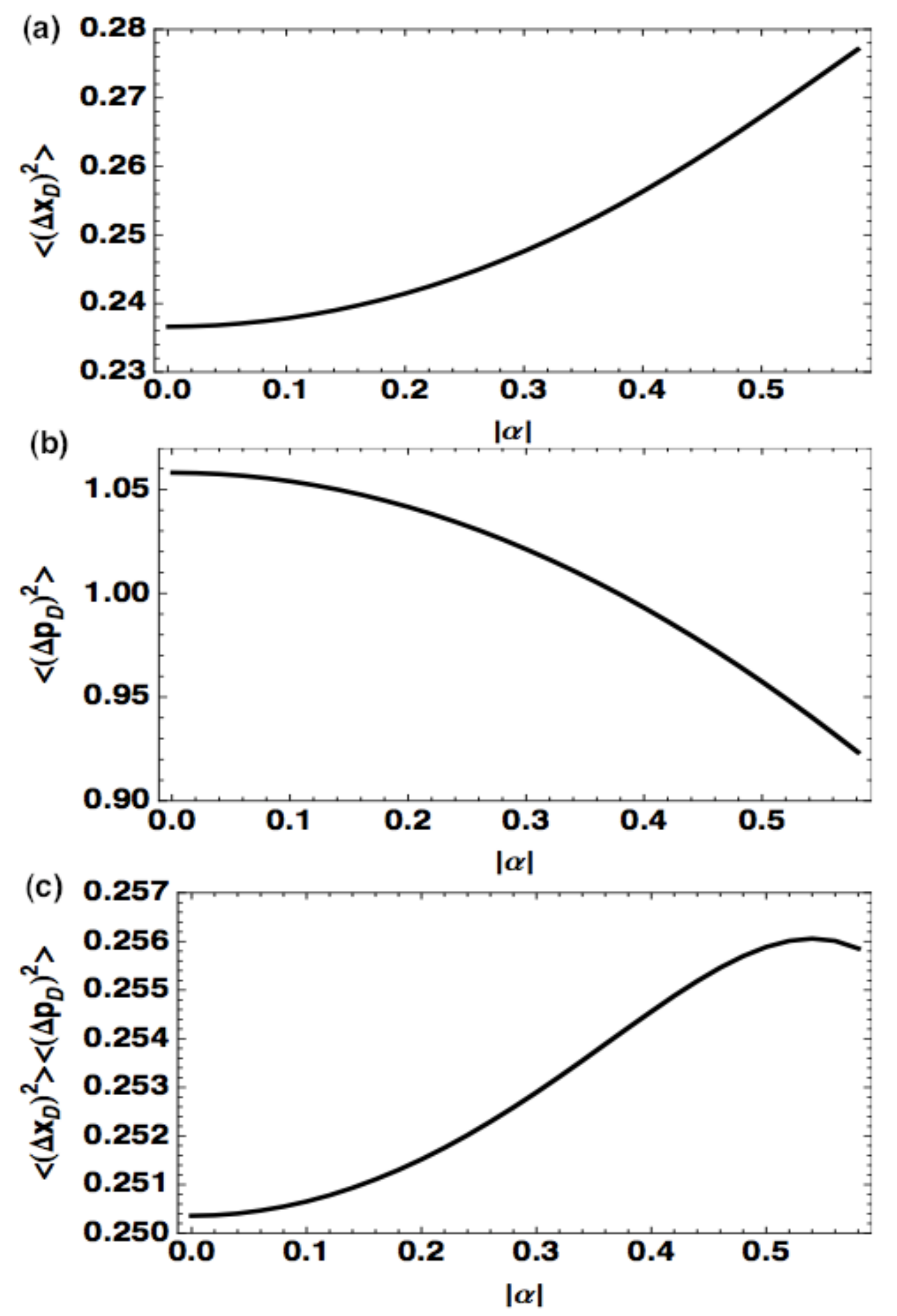} 
\end{center}
\caption{Average coordinate $\langle x_{D} \rangle$ (a), uncertainty in coordinate $\langle (\Delta x_{D})^{2} \rangle$ (b), and the corresponding uncertainty product $\langle (\Delta x_{D})^{2} \rangle \langle (\Delta p_{D})^{2} \rangle$ (c) at time $t=0$ as functions of $|\alpha|$ for the DOCS. In the calculation, a modified P\"oschl-Teller potential is modeled by a deformed oscillator with $s=10$ bound states.}
\label{fig:uncertainties-rm-alpha}
\end{figure}

Finally, in figure \ref{fig:uncertainties-rm-alpha} we see the dispersions in the deformed coordinate and momentum as a function of the parameter $|\alpha|$ evaluated at time $t=0$. Notice that for small $|\alpha|$ the dispersion in the deformed coordinate is less than 1/2 meaning squeezing, in contrast, that of the deformed momentum is larger than 1/2 and the uncertainty product is near the minimum possible value. As the parameter $|\alpha|$ increases the dispersion in the deformed coordinate increases so that the state is less squeezed and that of the momentum decreases but is still far from the minimum allowed value (in the range of values for the parameter $|\alpha|$ shown in the figures) so that the uncertainty product separates from the initial value. Because we are dealing with a finite number of bound states $s$ and we are not taking into account the states from the continuum we must use a small enough value of $|\alpha|$ so that the average occupation number is much smaller than $s$.
\medskip\\
To complete the examples, we just write the Displaced Operator Coherent States
associated we the trigonometric P\"oschl-Teller potential as
	\begin{equation}
	|\zeta(\alpha)\rangle=(1-|\zeta(\alpha)|^2)^\lambda\sum_{n=0}^\infty
	\zeta(\alpha)^n\sqrt{\frac{\Gamma(2\lambda+n)}{n!\Gamma(2\lambda)}}|n\rangle.
	\end{equation}
with $\zeta(\alpha)=e^{i\theta}\tanh(|\alpha|/\sqrt{2\lambda})$.
This expression should be compared with that of the AOCS (see Eq. \ref{AOCS-trigo}).

  \section{Discussion}
Based on the f-deformed oscillator formalism we have introduced non linear coherent states by generalization of two definitions, as eigenstates of a deformed annihilation operator AOCS and as the states that result by application of a deformed displacement operator on the vacuum state DOCS. We have applied our method to Hamiltonians that contain linear and quadratic terms in the number operator corresponding to the trigonometric and modified P\"oschl-Teller Hamiltonians and the Morse Hamiltonian. 

The AOCS and the DOCS obtained for the trigonometric P\"oschl-Teller potential are exact because the number of bound states stupported by the potential is infinite. On the other hand, the AOCS and the DOCS obtained for the Morse and the modified P\"oschl-Teller potentials are approximate because the number of bound states is finite. For the numerical results we have considered only the states obtained by application of a deformed displacement operator on the vacuum state, in Ref. \cite{ijtp08} we discussed the nonlinear coherent states obtained as eigenstates of a deformed annihilation operator for the Morse potential. Although from an algebraic-structure point of view the coherent states obtained from each generalization (DOCS) or (AOCS) are different, as well as their respective statistical behavior \cite{roman2}, the average values and the phase space trajectories obtained with them are almost identical \cite{santos11}. 

We must mentioned that the same form of nonlinear coherent states AOCS and DOCS
for the Morse and P\"oschl-Teller potentials would be obtained, if we had worked  
with the true ladder operators associated with each potential \cite{ricardo}. 
This happen because the ladder operators and deformed operators
share the same commutation relation for these kind of systems.
\medskip\\  
 {\bf Acknowledgements:}
 We thank Reyes Garc\'{\i}a for the maintenance of our computers and
 acknowledge partial support from CONACyT through project 166961 and DGAPA-UNAM project IN 108413.

\end{document}